\documentclass[graybox]{svmult}

\usepackage{type1cm}        
\usepackage{makeidx}         
\usepackage{graphicx}        
\usepackage{multicol}        
\usepackage[bottom]{footmisc}

\usepackage{subcaption}
 \usepackage{acronym}

\acrodef{adc}[ADC]{Analog-to-Digital Convertor}
\acrodef{dac}[DAC]{digital-to-analog convertor}
\acrodef{cs}[CS]{Compressed Sensing}
\acrodef{em}[EM]{ElectroMagnetic}
\acrodef{dtft}[DTFT]{discrete-time Fourier transform}
\acrodef{dnn}[DNN]{deep neural network} 
\acrodef{csi}[CSI]{Channel State Information}
\acrodef{map}[MAP]{maximum a-posteriori probability}
\acrodef{snr}[SNR]{Signal-to-Noise Ratio}
\acrodef{sinr}[SINR]{signal-to-interference-and-noise ratio}
\acrodef{bs}[BS]{Base Station} 
\acrodef{iot}[IOT]{Internet of Things}
\acrodef{mimo}[MIMO]{Multiple-Input Multiple-Output}
\acrodef{mse}[MSE]{Mean-Squared Error}
\acrodef{pdf}[PDF]{probability density function}
\acrodef{rv}[RV]{random variable}
\acrodef{tdd}[TDD]{time division duplexing}
\acrodef{rs}[RS]{Reed-Solomon}
\acrodef{lti}[LTI]{linear time-invariant}
\acrodef{wss}[WSS]{wide-sense stationary}
\acrodef{psd}[PSD]{power spectral density}
\acrodef{ser}[SER]{symbol error rate} 
\acrodef{ber}[BER]{bit error rate} 
\acrodef{isi}[ISI]{intersymbol interference}  
\acrodef{awgn}[AWGN]{additive white Gaussian noise} 
\acrodef{ut}[UT]{User Terminal} 
\acrodef{dc}[DC]{Direct Current} 
\acrodef{aoa}[AoA]{Angle of Arrival} 
\acrodef{mmw}[mmWave]{millimeter wave}
\acrodef{ris}[RIS]{Reconfigurable Intelligent Surface} 
\acrodef{hris}[HRIS]{Hybrid Reconfigurable Intelligent Surface} 
\acrodef{dma}[DMA]{Dynamic Metasurface Antenna}

\newcommand{\myVec}[1]{{\boldsymbol{#1}}}
\newcommand{\myMat}[1]{{\boldsymbol{#1}}}

\newcommand{\vecc}{{\operatorname{vec}}}
\usepackage{newtxtext}       %
\usepackage{newtxmath}       

\makeindex             

\begin{document}

\title*{Hybrid RISs for Simultaneous Tunable Reflections and Sensing }
\author{George C. Alexandropoulos, Nir Shlezinger, Ioannis Gavras, and Haiyang Zhang}
\institute{George C. Alexandropoulos \at Department of Informatics and Telecommunications, National and Kapodistrian University of Athens,  Greece, \email{alexandg@di.uoa.gr}
\and Nir Shlezinger \at School of Electrical and Computer Engineering, Ben-Gurion University, Israel, \email{nirshl@bgu.ac.il}
\and Ioannis Gavras \at Department of Informatics and Telecommunications, National and Kapodistrian University of Athens, Greece, \email{giannisgav@di.uoa.gr}
\and Haiyang Zhang \at School of Communication and Information Engineering, Nanjing University of Posts and Telecommunications, China,  \email{haiyang.zhang@njupt.edu.cn}
}
%
%
\maketitle

\abstract{The concept of smart wireless environments envisions dynamic programmable propagation of information-bearing signals through the deployment of Reconfigurable Intelligent Surfaces (RISs). Typical RIS implementations include metasurfaces with passive unit elements capable to reflect their incident waves in controllable ways. However, this solely reflective operation induces significant challenges in the RIS orchestration from the wireless network. For example,  channel estimation, which is essential for coherent RIS-empowered wireless communications, is quite challenging with the available solely reflecting RIS designs. This chapter reviews the emerging concept of Hybrid Reflecting and Sensing RISs (HRISs), which enables metasurfaces to reflect the impinging signal in a controllable manner, while simultaneously sensing a portion of it. The sensing capability of HRISs facilitates various network management functionalities, including channel parameter estimation and localization, while, most importantly, giving rise to computationally autonomous and self-configuring RISs. The implementation details of HRISs are first presented, which are then followed by a convenient mathematical model for characterizing their dual functionality. Then, two indicative applications of HRISs are discussed, one for simultaneous communications and sensing and another that showcases their usefulness for estimating the individual channels in the uplink of a multi-user HRIS-empowered communication system. For both of these applications, performance evaluation results are included validating the role of HRISs for sensing as well as integrated sensing and communications.}

\section{Introduction}
The potential of \acp{ris} for programmable \ac{em} wave propagation has recently motivated extensive academic and industrial interests, as a candidate smart connectivity paradigm for the sixth Generation (6G) of wireless communications \cite{Samsung,huang2019reconfigurable,di2019smart}. The \ac{ris} technology~\cite{ETSI_RIS_COMSTD}, which typically refers to artificial planar structures with almost passive electronic circuitry is envisioned to be jointly optimized with conventional transceivers~\cite{wu2021intelligent} in order to significantly boost wireless communications in terms of coverage, spectral and energy efficiency~\cite{9673796,10670007,9693982}, reliability~\cite{Samarakoon_2020_all,9406837}, and security~\cite{PLS_Kostas_all, Counteracting}, while satisfying regulated EM field emissions~\cite{RIS_challenges_all}.

The typical unit element of an RIS is the meta-atom, which is usually fabricated to realize multiple programmable states corresponding to distinct EM responses~\cite{RIS_challenges_all,10930892}. By externally controlling these states, various reflection and scattering profiles can be emulated \cite{Tsinghua_RIS_Tutorial}. The RIS hardware prototypes up to date do not include any power amplifying circuitry -this happens only in theoretical designs (e.g., \cite{9758764,9377648,10596064})-, hence, they mainly comprise metasurfaces that can only act as tunable reflectors. While passive \acp{ris} enable programmable wireless propagation environments, their purely reflective operation induces notable challenges when considered for wireless networking. For instance, 
the fact that \acp{ris} cannot sense the environment implies that they lack of information to dynamically configure their reflection pattern. For this reason, they are typically externally configured by a dedicated network entity or a \ac{bs} via dedicated control links\cite{10600711}, which in turn complicates their deployment and network-wise management \cite{9827873,RISE6G_COMMAG,10802983}. Furthermore, the inclusion of an \ac{ris} implies that a signal transmitted from a \ac{ut} to a \ac{bs} undergoes at least two channels: the \ac{ut}-\ac{ris} and \ac{ris}-\ac{bs} channels. Estimating these individual channels is a challenging task due to the reflective nature of \acp{ris} \cite{Tsinghua_RIS_Tutorial}, which significantly limits the ability to reliably communicate in a coherent manner. In addition, solely reflective \acp{ris} impose challenges on wireless localization \cite{wymeersch2020radio,RIS_Localization}. To overcome them, it was recently proposed to equip \acp{ris} with dedicated external devices comprising including reception Radio Frequency (RF) chains \cite{taha2021enabling}, enabling low-cost signal reception and relevant processing capabilities. Other recent designs, proposed interleaving purely reflecting elements with reception antenna arrays~\cite{HRIS_design_eucap2025}. 

Radiating metasurfaces have recently emerged as a promising technology for realizing low-cost, low-power, and large-scale \ac{mimo} antennas \cite{shlezinger2020dynamic,9847609,Gong_HMIMO_2023}. \acp{dma} pack large numbers of controllable radiative meta-atoms that are coupled to one or several waveguides, resulting in \ac{mimo} transceivers with advanced analog processing capabilities. While the implementation of \acp{dma} differs from reflective \acp{ris}, the similarity in their unit element structures indicates the feasibility of designing hybrid reflecting and sensing meta-atoms. Metasurfaces consisting of such meta-atoms can reflect their impinging signal, while simultaneously measuring a portion of it \cite{alamzadeh2021reconfigurable,HRIS_Mag_all}. Such Hybrid reflecting and sensing RISs (HRISs) bear the potential of significantly facilitating RIS orchestration, without notably affecting the coverage extension advantages offered by purely reflective RISs. 

In this chapter, the emerging concept of HRISs is reviewed together with its possible prominent applications for future wireless communications. The possible configurations for hybrid meta-atoms are first discussed, and then, an implementation of an HRIS consisting of simultaneously reflecting and sensing meta-atoms is presented. This is the focus of Section~2. To further investigate the potential of HRISs for wireless communications in Section~3, a mathematical model for HRIS-empowered wireless systems, in a manner that is amenable to system design, is provided. Two indicative applications of HRISs are discussed in Section~4, one for simultaneous communications and sensing, and another that showcases their usefulness for estimating the individual channels in the uplink. In the first application, the focus is on the joint optimization of the reconfigurable parameters of the BS and the HRIS with the goal to maximize the achievable downlink rate towards a single UT, while ensuring a minimum bistatic-type sensing accuracy over an Area of Interest (AoI) in the near-field region of the HRIS. The second application is an initial study on the potential gains of HRISs in multi-user MIMO communication systems, by considering the individual channels' estimation problem. The number of pilots needed to estimate the channels for the case without noise is first characterized, showcasing analytically the gains of HRISs compared to pure reflective \acp{ris}. Finally, Section~5 includes the concluding remarks of the chapter.

\section{HRIS Design and Proof-Of-Concept}
\label{sec:Hybrid}
A variety of RIS implementations has been presented \cite{RIS_challenges_all,Eucnc_Terrameta}, ranging from metasurfaces that manipulate wave propagation in rich scattering environments~\cite{RIS_Scattering_all} to improve the received signal strength, to those that realize desired anomalous reflection beyond Snell's law~\cite{WavePropTCCN}, in various frequency bands. In all those efforts, the RISs are incapable of any sort of sensing of the impinging signal. The concept of HRISs, referring to RISs realizing simultaneously sensing and communications, was first envisioned in \cite{alamzadeh2021reconfigurable}, where an implementation with sensing capabilities at the meta-atom level was demonstrated. HRISs incorporating this hybrid meta-atom design, while also leveraging previous works on DMAs \cite{shlezinger2020dynamic} to integrate sensing into each meta-atom, without losing their reconfigurable reflecting functionalities, are next presented.

\subsection{Hybrid Meta-Atoms} 
From a hardware perspective, one can propose two different configurations that allow for meta-atoms to both reflect and sense. The first consists of hybrid meta-atoms, which simultaneously reflect a portion of the impinging signal, while enabling the another portion to be sensed \cite{alamzadeh2021reconfigurable}. The second implementation uses meta-atoms that reconfigure between near-perfect absorption and reflection. In this case, the metasurface has two modalities \cite{ABSence}: it first senses the channel (absorption mode), and once necessary information is extracted, it enters the second mode (reflection) and directs the signal towards a desired direction. 

\begin{figure}[!t]
    \centering
    \includegraphics[width=\columnwidth]{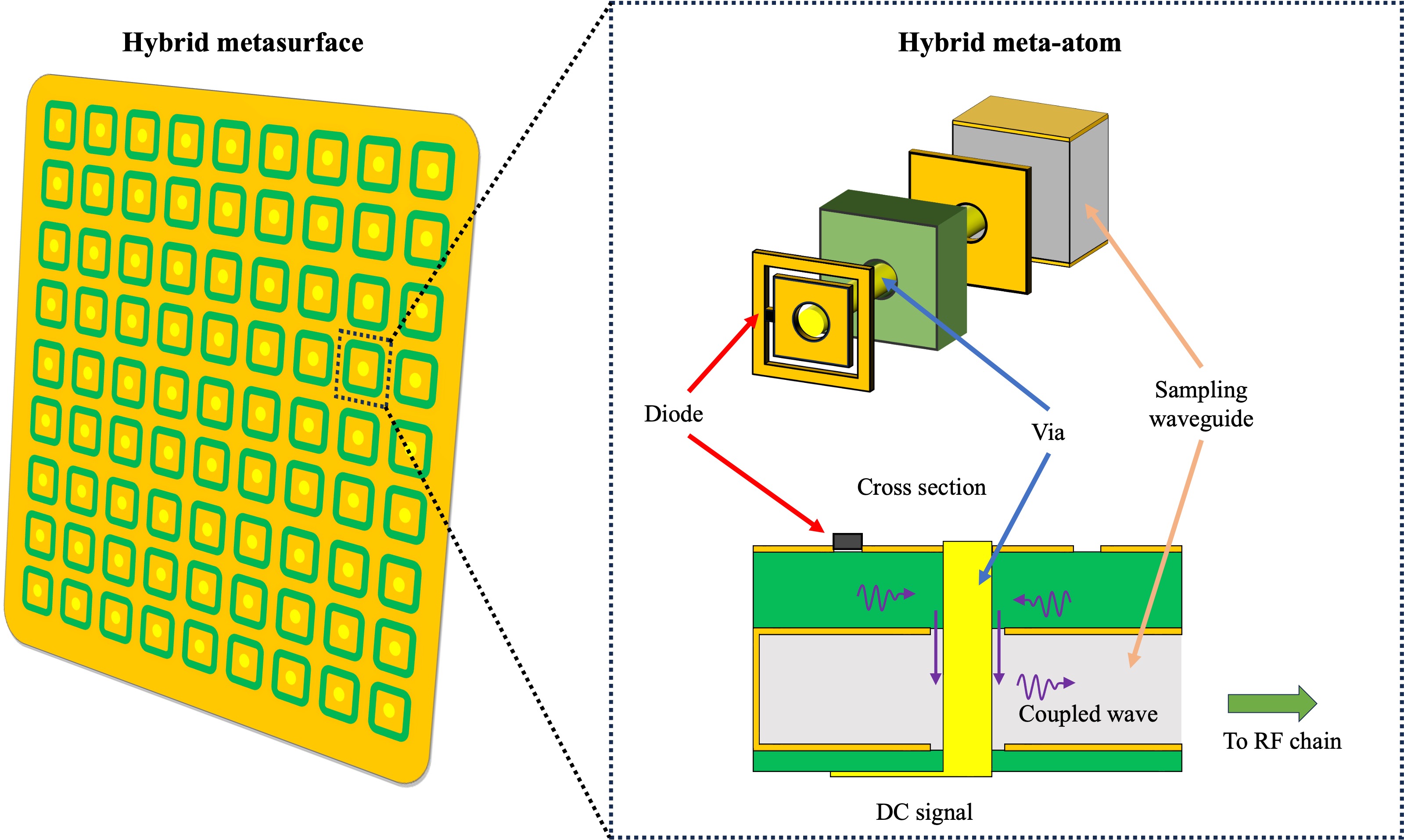}
    \caption{\small{Illustration of an HRIS and the constitutive hybrid meta-atom design proposed in \cite{alamzadeh2021reconfigurable}. The layers of the meta-atom on the top right are artificially separated for better visualization. The cross section of the metasurface, with a focus on a single meta-atom, and the coupled wave signal path are depicted on the bottom right drawing.}}
    \label{fig:HybridAtom_v01}
\end{figure}

Let us focus henceforth on the first case relying on hybrid meta-atoms, as illustrated in Fig.~\ref{fig:HybridAtom_v01}. Such HRISs are realized by adding a waveguide to couple to each or groups of meta-atoms. Each waveguide can be connected to a reception RF chain~\cite{hardware2020icassp}, allowing the HRIS to locally process a portion of the received signals in the digital domain via a dedicated baseband processor. However, elements' coupling to the waveguides implies that the incident wave is not perfectly reflected. In fact, the ratio of the reflected energy to the sensed one is determined by the coupling level. By keeping this waveguide near cutoff, its footprint can be reduced, while also reducing coupling to the sampling waveguide.

\subsection{Design Considerations} 
An important consideration in the HRIS design is the inter-element spacing. To accurately detect the phase front of the incident wave, the meta-atom spacing needs to be such that they are smaller than half a wavelength. However, this imposes constraints on the meta-atom size, which is accommodated via the multi-layer structure in Fig.~\ref{fig:HybridAtom_v01}. Another important factor to consider is the necessary circuitry to detect the signal coupled from each meta-atom; in particular, each waveguide should be connected to an RF chain. Since the incident wave on the HRIS may couple to all sampling waveguides (with different amplitudes), one can think of the combination of the metasurface and the sampling waveguides as a receiver structure with equivalent analog combining. 

The signal sensing/detection circuitry, however, adds another factor to consider when designing an HRIS: while closer element spacing improves the ability to direct the reflected signal to a desired direction (with smaller sidelobes) and increases the accuracy of detecting incident phase fronts, close element spacing increases the total number of meta-atoms, and consequently, the cost and potentially power consumption. Nonetheless, this can be balanced by limiting the number of RF chains---which are connected to the waveguides and not directly to each meta-atom--- and profiting from each meta-atom's controllable analog processing nature. Such a configuration also reduces the cost and power consumption of the introduced sensing circuitry, which does not appear in conventional \acp{ris} that are incapable of sensing. Recent approaches focus on reducing further the necessary detection circuitry, using, for example~\cite{10237986}, the intensity of the sampled incident signal to perform signal direction estimation. 

\subsection{Implementation}
Most unit elements in available RIS hardware designs are resonators, which can be easily modified to couple to a waveguide. For example, the meta-atom in \cite{sleasman2016microwave,boyarsky2021electronically} requires a via to deliver the DC signal to each element, in order to tune the switchable component. {In the hybrid meta-atom implementation proposed in \cite{alamzadeh2021reconfigurable},} a radial stub was added to the path of this conductive trace to diminish RF coupling. A Substrate Integrated Waveguide (SIW), which is effectively a rectangular waveguide, is used to capture the sampled wave, as shown in Fig.~ \ref{fig:HybridAtom_v01}. By changing the annular ring around the coupling via or the geometrical size of the SIW, the HRIS realizes different coupling strengths.

The hybrid meta-atom illustrated in Fig.~\ref{fig:HybridAtom_v01} is assumed to be
loaded by a varactor, whose effective capacitance is changed by an external Direct Current (DC) signal. The varying capacitance changes the phase of the reflected wave. By properly designing this phase variation along the HRIS, the reflected beam can be steered towards desired directions. To evaluate the ability of this HRIS design to simultaneously reflect while sensing its incident signal, a full-wave simulation study at $19$~GHz using Ansys HFSS was presented in~\cite{alamzadeh2021reconfigurable}. It was demonstrated that the designed hybrid meta-atom can be tuned to reflect and sense different portions of its impinging signal, while applying controllable phase shifting to the reflected signal. Additional studies of a full HRIS comprised of such hybrid meta-atoms, further showcasing the design's validity, can be found in~\cite{alamzadeh2021reconfigurable}.

\section{HRIS-Empowered Wireless Communications}
\label{sec:Comm}
In this section, the deployment of the HRIS concept, based on the implementation guidelines detailed in Section~\ref{sec:Hybrid}, for facilitating wireless communications is discussed. First, the envisioned HRIS-empowered wireless communication system is described. Next, a simple model that encapsulates the simultaneous tunable reflection and sensing operations of HRISs is presented. 
\begin{figure*}[!t]
    \centering
    \includegraphics[width=\linewidth]{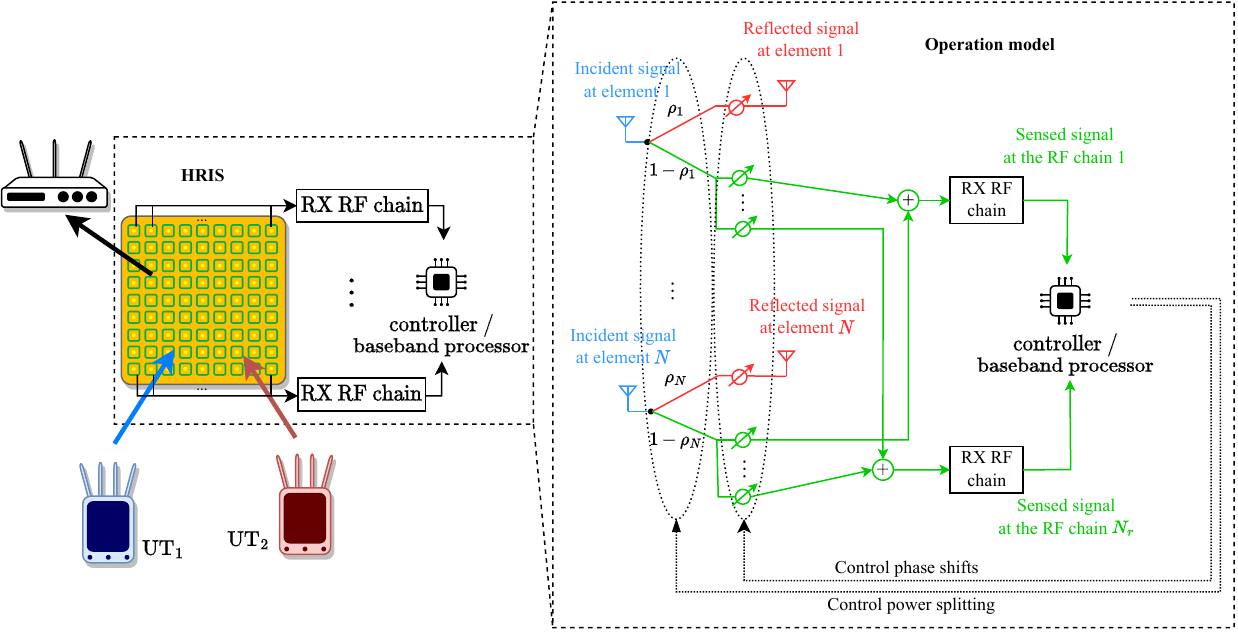}
    \caption{\small{The uplink of a two-user \ac{mimo} system incorporating the proposed HRIS that also senses a portion of the impinging UT signals, along with the operation model of the HRIS. In this model, the HRIS consists of $N$ hybrid meta-atoms. The incident signal at each meta-atom is split into a portion which is reflected (after tunable phase shifting), while the remainder of the signal is sensed and processed locally by a baseband processor hosted at the HRIS controller. Parameter $\rho_n$ ($n=1,2,\ldots,N$) represents the power splitting ratio of each $n$-th meta-atom. The received portion of the impinging signal undergoes the controllable element's response and is collected at the $N_r$ RF chains connected to the sampling waveguides, whose digital outputs are then processed by in the digital domain.}}
     \label{fig:SystemModel2}
\end{figure*}

\subsection{HRIS Mode of Operation}
\label{subsec:CommSystem}
The most common application of conventional \acp{ris} is to facilitate communication between UTs and the \ac{bs}~\cite{huang2019reconfigurable,9673796,10670007,9693982}, by shaping the \ac{em} signal propagation allowing to improve coverage and overcome harsh non-line-of-sight conditions. In such setups, the \ac{ris} tunes its unit elements to generate favorable propagation profiles. To achieve this~\cite{10600711,RISE6G_COMMAG,RIS_challenges}, the \ac{bs} maintains a control link with the digital controller of the \ac{ris}, where the latter sets the configuration of each element according to the control messages received by the \ac{bs}. This setup can be also extended to multiple \acp{ris} and cloud-controlled networks \cite{10670007,Samarakoon_2020_all,Alexandropoulos2022Pervasive,9410457}. This form of \ac{ris}-aided communications gives rise to several challenges. For instance, the fact that the \ac{ris} must be remotely controlled by the \ac{bs} complicates its deployment and network management, raising also security concerns~\cite{RIS_security_AIoT}. Moreover, in the absence of direct channels between the \ac{bs} and UTs, the former only observes the transmitted signals which propagated via the cascaded channels, namely, the composition of the \acp{ut}-\ac{ris} channel, the \ac{ris} reflection configuration, and the \ac{ris}-\ac{bs} channel. The fact that one cannot disentangle this combined channel implies that the \ac{bs} cannot estimate its individual components, but only the cascaded ones \cite{wang2020channel}. This property does not only reflect on the ability to estimate the channels, but also limits the utilization of some network management tasks, such as wireless localization and sensing. 

Consider an HRIS to be utilized for the same purposes as conventional reflective \acp{ris}. This includes, for instance, the typical application of coverage extension and connectivity boosting, by modifying the propagation profile of information-bearing \ac{em} waves, as illustrated in Fig.~\ref{fig:SystemModel2}. Similar to setups involving a conventional \ac{ris} \cite{huang2019reconfigurable,RIS_challenges}, the \ac{bs} can maintain a dedicated control link with the HRIS, though the latter can also operate independently due to its sensing capability~\cite{10802983}. For instance, the HRIS can use its sensed signal angle-of-arrival estimation to decide its phase configuration~\cite{HRIS_Mag_all}, thus, relieving its dependence on external configuration and control.
Nonetheless, when such a control link is present, it should not be used solely for unidirectional control messages from the \ac{bs} to the metasurface, as in reflective \acp{ris}, but now the HRIS can also use it to convey some sensed information to the \ac{bs}. 

The fact that the UTs' transmissions are also captured by the HRIS can notably improve channel estimation, as discussed in the sequel, which in turn facilitates coherent communications. Furthermore, the sensed signal can be used for localization~\cite{9726785,10124713,10557771,8313072}, sensing~\cite{HRIS_ISAC}, and RF mapping \cite{wymeersch2020radio,RFimaging}. As opposed to using reflective RISs with a few reception elements \cite{taha2021enabling}, HRISs can sense part of the signal during periodic acquisition phases, while fully reflecting during data transmission. An additional key difference between wireless networks empowered by HRISs and purely reflective \acp{ris} stems from the fact that HRISs do not reflect all the energy of the incident signal when it is also sensing, since a portion of it is sensed and collected by its local figital processor. Nevertheless, this may degrade the \ac{snr} at the \ac{bs}. 

The potential benefits of HRISs over conventional ones come with additional energy consumption and cost. While the proposed meta-atom design in Fig.~\ref{fig:HybridAtom_v01} is passive, as that of purely reflective \acp{ris}, HRISs require additional power for locally processing the sensed signals. The utilization of reception RF chain(s) and the additional analog combining circuitry, which are not required by conventional \acp{ris}, is translated into increased cost. While explicitly quantifying this increased cost and power consumption is highly subject to the specific implementation (see, e.g.,~\cite{GGM_ACCESS_2024} and reference therein), it is emphasized that an HRIS is still an \ac{ris}, and not a relay~\cite{Duong_relay_selection}. Namely, it does not involve a power-consuming wireless transmission mechanism for amplifying and transmitting its received signals, and is thus expected to maintain the cost and power gains of \acp{ris} over traditional relays \cite{huang2019reconfigurable,Duong_AF_relaying}. An additional challenge associated with HRISs stems from the optional need for a bi-directional control link between itself and the \ac{bs}. This link can be used for making HRIS's observations available to the BS (otherwise, they remain locally) to support higher throughput compared to the unidirectional control link utilized by conventional \acp{ris}.

\subsection{HRIS Model for Simultaneous Reflection and Sensing}
\label{subsec:Model}
\begin{figure}[!t]
		\centering			
		\includegraphics[width=1\columnwidth]{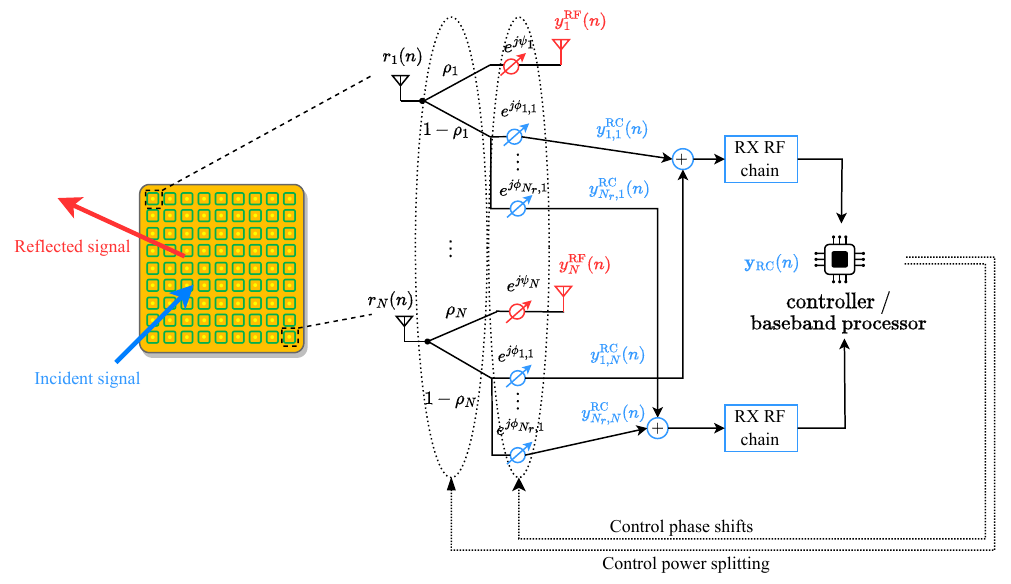}
		\vspace{-0.4cm}
		\caption{A simple model for the HRIS operation. The parameter $\rho_l$ models the portion of the impinging signal at the $l$-th meta-atom of the HRIS that gets tunably reflected, while $\psi_l$ and $\phi_{r,l}$ model respectively the meta-atom's controllable phase shift and the joint effect of its response together with the analog phase shift before being guided into the $r$-th receive RF chain.} 
		\label{fig:Hybrid_RIS_model}
	\end{figure}

Herein, a simple, yet convenient, model for the HRIS operation is presented. To model the dual reflection-reception functionality of HRISs, a hybrid metasurface comprised of $N$ meta-atom elements is considered, where each element is capable of simultaneously reflecting a portion of the impinging signal, while simultaneously waveguiding another portion of it to the local digital processor in a controllable manner. Let $r_{l}(n)$ denote the radiation observed by the $l$-th HRIS element ($l=1,2,\ldots,N$) at the $n$-th time instance. A portion of this signal, dictated by the parameter $\rho_l \left( n \right) \in [0,1]$, is reflected with a controllable phase shift $\psi_l \left( n \right) \in [0,2\pi)$, and thus the reflected signal from the $l$-th element at the $n$-th time instant can be mathematically expressed as follows:
\begin{equation}\label{eq:reflection}
y_l^{\rm RF} (n)= \rho_l\left( n \right)e^{\jmath \psi_l \left( n \right)} r_l(n). 
\end{equation} 
The remainder of the observed signal is locally processed via analog combining and then digital processing. The signal forwarded to the $r$-th RF chain ($r=1,2,\ldots,N_r$) via combining, from the $l$-th element at the $n$-th time instant is consequently given by the expression:
\begin{equation}\label{eq:power_split_t}
y_{r,l}^{\rm RC} (n) = (1 - \rho_l \left( n \right))e^{\jmath \phi_{r,l} \left( n \right)} r_l(n), 
\end{equation}
where $\phi_{r,l} \left( n \right)\in[0,2\pi)$ represents the adjustable phase that models the joint effect of the response of the $l$-th meta-atom and the subsequent analog phase shifting. The aforedescribed HRIS operation model is illustrated in Fig.~\ref{fig:Hybrid_RIS_model}.

It is noted that the operation of conventional passive and reflective RISs can be treated as a special case of the HRIS architecture, by setting all $\rho_l \left( n \right)$ in \eqref{eq:reflection} equal to $1$. 
Compared with existing relay techniques~\cite{DF_relays}, HRISs bring forth two major advantages. First, HRISs allow full-duplex operation (i.e., simultaneous reflection and reception) without inducing any self interference, which is unavoidable in full-duplex relaying systems~\cite{7997175,9933358}. Second, HRISs require low power consumption since they do not need power amplifiers utilized by active transmit arrays; a typical receive RF consists of a low noise amplifier, a mixer that downconverts the signal from RF to baseband, and an Analog-to-Digital Converter (ADC). Note that harvesting solutions for empowering HRISs were proposed in~\cite{10447070}, while~\cite{GGM_ACCESS_2024} discussed the implementation of the reception RF chains of a receiving DMA, and thus similarly an HRIS, with very basic ADCs of $1$-bit resolution. HRISs are also different from the concept of active RISs, which integrate the power amplifiers into the antenna elements of RISs~\cite{zhang2022active,mine_Active_RIS}. Hence, active RISs are capable of amplifying the reflected signals, but without the signal reception and decoding capability.
	
The resulting signal model at the HRIS can be expressed in vector form in baseband representation, as follows. By stacking the received signals $r_l(n)$, with $l=1,2,\ldots,N$, and the reflected signals $y_l^{\rm RF}(n)$, with $l=1,2,\ldots,N$, at the $N\times 1$ complex-valued vectors $\mathbf{R}(n)$ and  $\mathbf{y}_{\rm RF}(n)$, respectively, it follows from \eqref{eq:reflection} that:
\begin{equation} \label{eq:reflection_vector}
    \mathbf{y}_{\rm RF}(n) = \boldsymbol{\Psi} \left(\boldsymbol{\rho}\left(n\right), \boldsymbol{\psi}\left(n\right)\right) \mathbf{R}(n)
\end{equation}
with 
\begin{equation} \label{eq: reflection_matrix}
    \myMat{\Psi } \left( \boldsymbol \rho  \left( n \right), \boldsymbol \psi  \left( n \right)\right)  \triangleq {\rm diag} \left(\left[\rho_1  \left( n \right)\,e^{\jmath \psi_1  \left( n \right)},\rho_2  \left( n \right)\,e^{\jmath \psi_2  \left( n \right)},\\ \dots, \rho_N  \left( n \right)\,e^{\jmath \psi_N  \left( n \right)}\right]\right).
\end{equation}
Similarly, by letting $\mathbf{y}_{\rm RC}(n)\in \mathbb{C}^{N_r\times1}$ be the reception output vector at the HRIS, the following expression is deduced (neglecting for now the contribution of the recpetion thermal noise):
\begin{equation} \label{eq:output_compact}
    \mathbf{y}_{\rm RC}(n) = {\boldsymbol \Phi}\left( \boldsymbol \rho  \left( n \right), \boldsymbol \phi  \left( n \right)\right) \mathbf{R}(n),
\end{equation}
where the $N_r \times N$ matrix $\myMat{\Phi}\left( \boldsymbol \rho  \left( n \right), \boldsymbol \phi  \left( n \right)\right)$ represents the analog combining carried out at the HRIS receiver. When the $l$-th meta-atom element is connected to the $r$-th reception RF chain, then $[\myMat{\Phi}\left( \boldsymbol \rho  \left( n \right), \boldsymbol \phi  \left( n \right)\right)]_{r,l} =  (1 - \rho_l \left( n \right))e^{\jmath \phi_{r,l} \left( n \right)}$ $\forall r,l$, while, when there is no such connection (e.g., for partially-connected analog combiners), it holds that
$[\myMat{\Phi}\left( \boldsymbol \rho  \left( n \right), \boldsymbol \phi  \left( n \right)\right)]_{r,l} = 0$ $\forall r,l$. 

The reconfigurability of HRISs implies that the parameters $\boldsymbol \rho  \left( n \right)$ as well as the phase shifts $\boldsymbol \psi  \left( n \right)$ and $\boldsymbol \phi  \left( n \right)$ are externally controllable. It is noted that when an element is connected to multiple reception RF chains, then additional dedicated analog circuitry (e.g., conventional networks of phase shifters) is required to allow the signal to be forwarded with a different phase shift to RF chain; this, of course, at the possible cost of additional power consumption. Nonetheless, when each element feeds a single RF chain, then the model in Fig.~\ref{fig:Hybrid_RIS_model} can be realized without such circuitry by placing the elements on top of separated waveguides (see, e.g., \cite{shlezinger2020dynamic}).

\section{HRIS Optimization for Indicative Applications}
\label{subsec:Estimation}
In this section, the contribution of the sensing capabilities of HRISs in two indicative applications are investigated: simultaneous communications and sensing, as well as estimation of the end-to-end channel matrices in uplink multi-user communication systems. The former is key to enabling self-configuring HRISs, alleviating their dependence on external control, and thus, notably facilitating deployment compared with conventional \acp{ris}. This application also highlights the deployment of HRISs for Integrated Sensing And Communications (ISAC)~\cite{RIS_ISAC_SPM}. The second applications showcases how, by sensing a portion of its impinging signal, HRISs can contribute to estimating the composite \acp{ut}-HRIS channel and the HRIS-\ac{bs} channel (see Fig.~\ref{fig:SystemModel2}). Such estimation is highly challenging when using a purely reflective \ac{ris}, as one only observes the channel outputs at the \ac{bs} \cite{hu2019two,taha2021enabling},  and the common practice is to resort to estimating the combined channel effect, i.e., the cascaded channel~\cite{wang2020channel,LZA_TWC_2021,LHA_TCOM_2021,LHG_TWC_2022,Tsinghua_RIS_Tutorial,BGA_TWC_2024}, using pilot reflective patterns at the RIS. 

\subsection{Simultaneous Communications and Sensing}
Consider a wireless ISAC system setup consisting of an extremely large MIMO BS, an HRIS, and a single-antenna UT. The BS wishes to communicate in the downlink direction with the UT via the assistance of the HRIS, which is also tasked to enable sensing over an AoI within its vicinity that includes $K$ passive radar targets. This sensing functionality takes place on a bistatic manner via the received reflected downlink signals from the targets at the HRIS side. Without loss of generality, it is assumed that both the UT and all $K$ radar targets are located within the near-field region of the HRIS, while this metasurface may or may not be in the near field of the BS. 

\subsubsection{System Model}
The BS node is equipped with a Uniform Planar Array (UPA) consisting of $M\triangleq M_{\rm R}\times M_{\rm E}$ antennas, and is capable of realizing fully digital transmit BF. It is assumed that the BS lies at the origin in the $xz$-plane. The HRIS is also modeled as a UPA with $N\triangleq N_{\rm RF}N_{\rm E}$ elements positioned in the $xz$-plane opposite of the BS, with its first element lying at the point $(r_{\rm RIS},\theta_{\rm RIS},\varphi_{\rm RIS})$ representing respectively the distance from the origin as well as the elevation and azimuth angles. The hybrid meta-atoms at the each column of the HRIS are assumed attached via a dedicated waveguide to a single reception RF chain (there exist in total $N_{\rm RF}$ with $N/N_{\rm RF}\in\mathbb{Z}^+_*$), enabling the absorption of a certain portion of the impinging signals for baseband processing at the HRIS controller. For the sake of simplicity, it is assumed that the spacing between adjacent vertical and horizontal elements at both the BS and HRIS UPAs is $\lambda/2$, where $\lambda$ is the wavelength of the communications signal. Finally, the common assumption that the BS and the HRIS are aware of each other's static positions is adopted, as well as that the latter's controller shares its signal observations as well as its reflection configurations with the former, via a reliable link, which is responsible for executing the targeted optimization.

The simultaneous reflective and sensing functionality of the HRIS is controlled through $N$ identical power splitters (thus, a common power splitting ratio, i.e., $\rho_\ell=\rho$ $\forall l$, is assumed for all HRIS elements), which divide the power of the impinging signal at each hybrid meta-atom in the respective parts. For the sensing operation, to feed the absorbed portion of the impinging signal to the $N_{\rm RF}$ reception RF chains, the HRIS applies the partially-connected analog combining matrix $\mathbf{\Phi}(\rho,\boldsymbol{\phi})\in\mathbb{C}^{N_{\rm RF} \times N}$, which is modeled as (neglecting the time index $n$ in this section):
\begin{align}
    [\mathbf{\Phi}(\rho,\boldsymbol{\phi})]_{j,(r-1)N_{\rm E}+l} = \begin{cases}
    (1-\rho)e^{\jmath\phi_{r,l}},&  r=j\\
    0,              & r\neq j
\end{cases},
\end{align}
with $l=1,2,\ldots,N$ and $r=1,2,\ldots,N_{\rm RF}$, for each non-zero element in $\mathbf{\Phi}$. In addition, the HRIS reflection configuration is represented by $\boldsymbol{\Psi}(\rho,\boldsymbol{\psi})=\rho\left[ e^{\jmath \psi_1 }, e^{\jmath \psi_2},\\ \dots, e^{\jmath \psi_N}\right]\in\mathbb{C}^{N\times N}$ similarly defined as in \eqref{eq: reflection_matrix}. Finally, the BS precodes digitally via the channel-dependent BF vector $\mathbf{v}\in\mathbb{C}^{M\times 1}$ each complex-valued symbol $s$ such that $\mathbb{E}\{\|\mathbf{v} s\|^2\}\leq P_{\rm max}$, where $P_{\rm max}$ represents the maximum transmission power. The baseband received signal $\mathbf{y}_{\rm RC}\in\mathbb{C}^{N_{\rm RF}\times 1}$ at the outputs of the $N_{\rm RF}$ RX RF chains of the HRIS can be mathematical expressed, similarly to \eqref{eq:output_compact}, as:
\begin{align}\label{eq: y_ref}
    \mathbf{y}_{\rm RC} \triangleq \mathbf{\Phi}(\rho,\boldsymbol{\phi})\mathbf{H}_{\rm RB}\mathbf{v}\bar{s}+\mathbf{n},
\end{align}
where $\mathbf{H}_{\rm RB} \in \mathbb{C}^{N \times M}$ represents the composite channel matrix between the HRIS and the BS, which also depends on the location of the $K$ targets and the UT, $\bar{s}\in\mathbb{C}$ corresponds to the transmitted symbol, and $\mathbf{n} \sim \mathcal{CN}(0, \sigma^2 \mathbf{I}_{N_{\rm RF}})$ represents the AWGN vector. It is important to note that, in \eqref{eq: y_ref}, the contribution of the static HRIS-BS channel in $\mathbf{H}_{\rm RB}$ (i.e., excluding the reflections from the targets and the UT) can be entirely eliminated, leveraging the reasonable assumption that the HRIS has prior knowledge of the BS's position.

\subsubsection{Problem Formulation and Solution}
Let us now focus on the joint optimization of the reconfigurable parameters of the BS and the HRIS with the goal to maximize the achievable downlink rate towards the single UT, while ensuring a minimum bistatic-type sensing accuracy over an AoI in the vicinity of the HRIS. For the latter requirement, the ISAC system's ability to maintain a consistent level of estimation accuracy for the target spatial parameters across the entire AoI is considered. To this end, the AoI is discretized into a finite set of points, each represented by the polar coordinates $\boldsymbol{\zeta}_q\triangleq [r_q, \varphi_q, \theta_q]\in\mathbb{R}^{1\times 3}$ $\forall q=1,2,\ldots,\mathcal{Q}$. It is evident from \eqref{eq: y_ref}'s inspection that, for a coherent channel block involving $\Bar{T}$ unit-powered symbol transmissions, with $\bar{\mathbf{s}}=[\bar{s}(1),\bar{s}(2),\ldots,\bar{s}(\Bar{T})]\in\mathbb{C}^{1\times \Bar{T}}$, yields $\Bar{T}^{-1}\mathbf{s}\mathbf{s}^{\rm H}=1$, and the received signal at the output of the HRIS's RX RF chains is distributed as $\mathbf{y}_{\rm RC}\sim\mathcal{CN}(\mathbf{M},\mathbf{C}_n)$, with mean $\mathbf{M} \triangleq \mathbf{\Phi}(\rho,\boldsymbol{\phi})\mathbf{H}_{\rm RB}\mathbf{v}\bar{\mathbf{s}}$ and covariance matrix $\mathbf{C}_n\triangleq\sigma^2\mathbf{I}_{N_{\rm RF}}$. In the context of area-wide sensing over the set of spatial points $\boldsymbol{\xi} = [\boldsymbol{\zeta}_1, \boldsymbol{\zeta}_2, \ldots, \boldsymbol{\zeta}_Q] \in \mathbb{R}^{1 \times 3Q}$, and assuming that $\mathbf{M}$ is computed with respect only to the positions within $\boldsymbol{\xi}$, each $(i,j)$-th element (with $i,j=1,2,\ldots,3Q$) of the $3Q\times3Q$ Fisher Information Matrix (FIM) $\mathbf{J}$ can be calculated as follows:
\begin{align}
    \nonumber[\mathbf{J}]_{i,j} \!\triangleq\! 2\Re\left\{\!\frac{\partial \mathbf{M}^{\rm H}}{\partial[\boldsymbol{\xi}]_i}\mathbf{C}_n^{-1}\frac{\partial \mathbf{M}}{\partial[\boldsymbol{\xi}]_j}\!\right\}\!+\!\text{Tr}\left\{\!\mathbf{C}_n^{-1}\frac{\partial\mathbf{C}_n}{\partial[\boldsymbol{\xi}]_i}\mathbf{C}_n^{-1}\frac{\partial\mathbf{C}_n}{\partial[\boldsymbol{\xi}]_j}\!\right\}\!.
\end{align}
Since $\mathbf{C}_n$ is independent of $\boldsymbol{\xi}$, it holds that $\nabla_{\boldsymbol{\xi}}\mathbf{C}_n = \mathbf{0}_{1\times3Q}$. Therefore, each element of the FIM depends solely on the mean, and $\forall i$ the following result can be easily deduced: $\frac{\partial\mathbf{M}}{\partial[\boldsymbol{\xi}]_i} = \mathbf{\Phi}\frac{\partial\mathbf{H}_{\rm RB}}{\partial[\boldsymbol{\xi}]_i}\mathbf{v}\mathbf{s}$. Consequently, the $(i,j)$-th diagonal  of the FIM can be re-written as follows:
\begin{align}
    [\mathbf{J}]_{i,j} = \frac{2\rho^2\Bar{T}}{\sigma^2}\Re\Bigg\{\mathbf{v}^{\rm H}\frac{\partial\mathbf{H}_{\rm RB}^{\rm H}}{\partial[\boldsymbol{\xi}]_i}\mathbf{\Phi}\mathbf{\Phi}^{\rm H}\frac{\partial\mathbf{H}_{\rm RB}}{\partial[\boldsymbol{\xi}]_j}\mathbf{v}\Bigg\}.\nonumber
\end{align}
Combining all above together, the Position Error Bound (PEB) for the discretized AoI at positions $\boldsymbol{\xi}$ can be expressed as:
\begin{align}\label{eq: PEB}
\text{PEB}_{\boldsymbol{\xi}} \triangleq \sqrt{{\rm CRB}_{\boldsymbol{\xi}}}=
\sqrt{{\rm Tr}\left\{\mathbf{J}^{-1}\right\}}. 
\end{align}


Based on the aforementioned PEB analysis, the herein considered HRIS-enabled ISAC design objective is expressed as follows:
\begin{align}
        \mathcal{OP}\!:& \nonumber\underset{\substack{\mathbf{\Phi},\mathbf{\Psi},\mathbf{v}}}{\max} \,\,\log_2\left(1+\sigma^{-2}\|\widehat{\mathbf{h}}_{\rm DL}\mathbf{v}\|^2\right)\\
        &\nonumber\text{s.t.}\, {\rm PEB}_{\boldsymbol{\xi}}\leq\gamma_s\,, \left|\left[\mathbf{\Phi}\right]_{r,l}\right|=1,\, \left|\left[\mathbf{\Psi}\right]_{l,l}\right|=1\,\forall r,l \,\left\|\mathbf{v}\right\|^2\leq P_{\rm max}.
\end{align} 
In this formulation, $\gamma_s$ indicates the desired localization accuracy across the AoI for the ${\rm PEB}_{\boldsymbol{\xi}}$. 
In addition, the vector $\widehat{\mathbf{h}}_{\rm DL}\triangleq\widehat{\mathbf{h}}_{\rm UB}+(1-\rho)\widehat{\mathbf{h}}_{\rm UR}\mathbf{\Psi}\widehat{\mathbf{H}}_{\rm RB}\in\mathbb{C}^{1\times M}$ denotes the estimation of the downlink channel towards the UT, using the estimations $\widehat{\mathbf{h}}_{\rm BU}\in\mathbb{C}^{1\times M}$, $\widehat{\mathbf{h}}_{\rm RU}\in\mathbb{C}^{1\times N}$, and $\widehat{\mathbf{H}}_{\rm RB}\in\mathbb{C}^{N\times M}$ for $\mathbf{h}_{\rm BU}$, $\mathbf{h}_{\rm RU}$, and $\mathbf{H}_{\rm RB}$, which represent the channel gain matrices between the UT and the BS, the UT and the HRIS, and the HRIS and the BS, respectively. Note, that the considered channels can be obtained via typical channel estimation schemes~\cite{alexandropoulos2020hardware}. Due to the coupled and non-convex nature of $\mathcal{OP}$, this ISAC design problem is solved using an alternating optimization approach. Specifically, we first optimize the analog and digital beamformers, $\mathbf{\Phi}$ and $\mathbf{v}$, via semidefinite relaxation, and then optimize the reflection phase shift matrix, $\mathbf{\Psi}$, using simulated annealing. This process takes place in an iterative manner until a termination criterion is satisfied.


\subsubsection{Numerical Results}
\begin{figure}[!t]
    \centering
    \includegraphics[width=\columnwidth]{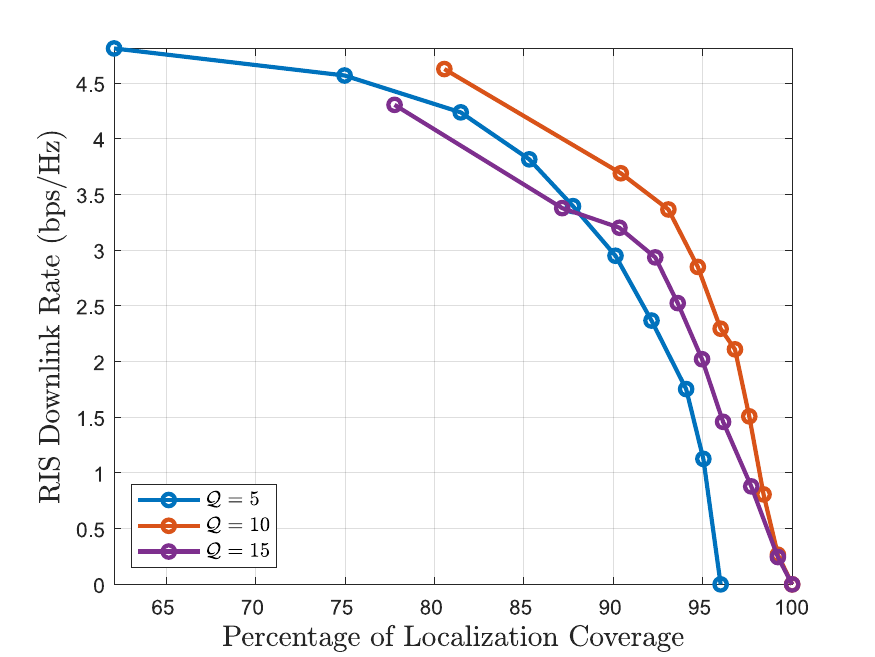}
    \caption{\small{The trade-off between achievable downlink rate and localization coverage performance for a setup including an HRIS with $N=256$ meta-atoms and a BS with $M=16$ antenna elements, considering $\Bar{T}=100$ symbol transmissions.}}
    \label{fig:tradeoff}
\end{figure}

In the following analysis, we numerically evaluate the simultaneous communications and localization performance of the proposed HRIS-enabled ISAC framework. We have simulated a narrowband setup with bandwidth $B = 150$ kHz centered at the frequency of $28$ GHz, where coherent channel blocks span $\Bar{T}=100$ transmissions. The AoI was defined at the fixed elevation angle of $\theta = 30^\circ$, azimuth angle of $\varphi \in [20^\circ, 80^\circ]$, and range of $r\in[1, 15]$ meters. Both the UE and the $K=2$ radar targets were randomly positioned within the AoI, while the HRIS was considered to be placed at the point with $\theta_{\rm RIS} = 30^\circ$, $\varphi_{\rm RIS} = 60^\circ$, and $r_{\rm RIS} = 16$ meters. We have used $500$ Monte Carlo trials for the following simulation results, considering a BS equipped with a $2 \times 8$ BS UPA and an HRIS with $N_{\rm RF} = 4$ RX RF chains each connected to $N_{\rm E} = 64$ hybrid meta-atoms. We have set AWGN's variance as $\sigma^2 = -174 + 10\log_{10}(B)$ and in order to evaluate the  coverage performance of the proposed system, we compute the localization coverage as the percentage of discrete points within the AoI-sampled at $1000$ equidistant locations-that satisfy PEB performance lower than $\gamma_s$.

Figure~\ref{fig:tradeoff} illustrates the trade-off between the achievable DL rate and the localization coverage performance for the proposed HRIS-enabled ISAC system design, operating with a maximum transmission power $P_{\rm \max}=16$ dBm, $\gamma_s=10^{-3}$ and  for $10$ different absorption levels $\rho$, evenly distributed within the range $(0,1]$ (the 'o' marks), and for varying numbers of discretization points, $\mathcal{Q} = \{5, 10, 15\}$ of the AoI. As expected, increasing the absorption level leads to improved localization coverage at the expense of degraded communication performance. Additionally, the results show that enhanced localization coverage can still be achieved at lower absorption levels by increasing the discretization resolution of the AoI. In essence, increasing the number of sampling points within the AoI leads to a greater allocation of resources for sensing, as the system must account for more directions, resulting in the design of broader sensing beams. Moreover, focusing on a larger number of directions introduces a cooperative effect among adjacent points, as minimizing the PEB at one location can indirectly contribute to reducing the PEB at neighboring locations. However, this improvement plateaus as $\mathcal{Q}$, approaches the limit set by the available beamforming degrees of freedom, i.e., $\min(N, M)$ beyond which performance begins to degrade. This limitation arises from the system’s inability to form sharp beams in all desired directions simultaneously when the number of focusing directions is equal or exceeds the number of available beamforming degrees of freedom (e.g., 15 directions for sensing and 1 for communication), leading to reduced spatial resolution and, consequently, diminished sensing accuracy.


\subsection{Estimation of the Individual Channels}
To demonstrate the channel estimation capability of H\acp{ris}, the uplink system of Fig.~\ref{fig:SystemModel2} with $K$ \acp{ut} and an H\ac{ris} with $N$ meta-atoms is considered in this section. As before, the HRIS has $N_r$ RF chains (which can be greater than one), and maintains a bidirectional control link with the \ac{bs}~\cite{RIS_challenges_all,10600711}, as in Fig.~\ref{fig:SystemModel2}. A simple strategy to estimate the individual channels is to have the H\ac{ris} estimate the HRIS-\acp{ut} channel based on its sensed observations, and forward this estimate to the \ac{bs} over the control link (wired or wireless), while changing the phase configuration between pilot symbols as in \cite{wang2020channel,Tsinghua_RIS_Tutorial}. The fact that the H\ac{ris} also reflects while sensing, allows the \ac{bs} to estimate the \ac{bs}-H\ac{ris} channel from its observed reflections. This can happen by effectively re-using the transmitted pilots for estimating both the HRIS-\acp{ut} and \ac{bs}-H\ac{ris} channels. 

\subsubsection{System Model}
Consider an uplink multi-user \ac{mimo} system where a \ac{bs} equipped with $M$ antenna elements serves $K$ single antenna \acp{ut} with the assistance of an HRIS, as illustrated in Fig.~\ref{fig:SystemModel2}. Let us assume that there is no direct link between the \ac{bs} and any of the $K$ \acp{ut}, thus, communication is exclusively enabled only via the HRIS. Let $\mathbf{H}_{\rm RB} \in \mathbb{C}^{M \times N}$ denote the channel gain matrix between the \ac{bs} and HRIS, and $\mathbf{h}_{{\rm RU},k} \in \mathbb{C}^{N}$ be the channel gain vector between the HRIS and each of the $k$-th UT ($k=1,2,\ldots,K$). Without loss of generality, consider independent and identically distributed (i.i.d.) Rayleigh fading for all channels, implying that with $\mathbf{H}_{\rm BR}$ and each $\mathbf{h}_{{\rm RU},k}$ have i.i.d. zero-mean Gaussian entries with variances $\beta$ and $g_k$, respectively, denoting the path losses. In addition, for notation simplicity, the matrix definition ${\mathbf{G}}\triangleq\left[\mathbf{h}_{{\rm RU},1},\mathbf{h}_{{\rm RU},2},\ldots,\mathbf{h}_{{\rm RU},K}\right]$ is used.
 
 \begin{figure}
    \centering
    \includegraphics[width=0.8\columnwidth]{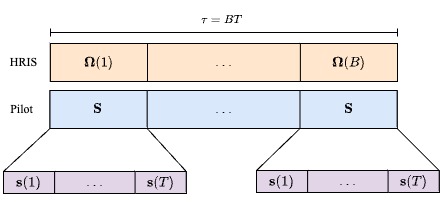}
    \caption{The frame structure for estimating the individual HRIS-\acp{ut} and \ac{bs}-H\ac{ris} channels in the uplink HRIS-empowered multi-user \ac{mimo} system of Fig.~\ref{fig:SystemModel2}.}
    \label{fig:structure}
\end{figure}

A simple pilot-based channel training protocol is considered, where the channel estimation time $\tau$ is divided into $B$ sub-frames, and each sub-frame consists of $T$ times slots such that $\tau = BT$, as depicted in Fig.~\ref{fig:structure}. It is assumed that the channel coherence time is larger than the channel estimation time, i.e., the duration of the $B$ sub-frames. The reconfigurable parameters of the HRIS remain constant during each sub-frame of $T$ time slots and vary from one sub-frame to another. Orthogonal pilot sequences $\left\{\mathbf{s}_k \right\}_{k=1}^K$ are sent repeatedly over the $B$ sub-frames, where $\mathbf{s}_k\triangleq\left[s_k\left(1\right),s_k\left(2\right),\ldots,s_k\left( T\right)\right] \in \mathbb{C}^{1 \times T}$ is the pilot sequence of the $k$-th \ac{ut} satisfying for $1 \leq k_1,k_2 \leq K$: $\mathbf{s}_{k_1}\mathbf{s}_{k_2}^{\rm H} = T$, if $k_1 = k_2$; and $\mathbf{s}_{k_1}\mathbf{s}_{k_2}^{\rm H} = 0$, if $k_1 \neq k_2$. In Fig.~\ref{fig:structure}, $\mathbf{s}(t) \triangleq \left[s_1\left( t\right),s_2\left( t\right),\ldots,s_K\left( t\right)\right]^{\rm T}$ collects the pilot signals of the $K$ \acp{ut} at each $t$-th time slot for each sub-frame, and ${\boldsymbol \Omega}\left(b\right)\triangleq \left[{\boldsymbol \rho}(b), {\boldsymbol \phi}(b),{\boldsymbol \psi}(b)\right] $ includes all the optimization variables of the HRIS at each $b$-th sub-frame. Consequently, the  signal received at the HRIS at each $t$-th time slot for each $b$-th sub-frame is given by:
 \begin{equation}\label{eq:received_all_pilot_output_new}
    \mathbf{y}_{\rm RC}\left( b,t \right)={\boldsymbol \Phi}\left( {\boldsymbol \rho}(b), {\boldsymbol \phi}(b)\right) \mathbf{G} \mathbf{s}(t) + \mathbf{z}_{\rm RC}\left( b,t \right), 
\end{equation}
where ${\boldsymbol \Phi}\left( {\boldsymbol \rho}(b), {\boldsymbol \phi}(b)\right)$ represents the reception matrix of the HRIS during the $b$-th sub-frame and $\mathbf{z}_{\rm RC}\left( b,t \right) \in\mathbb{C}^{N_r}$ is a zero-mean \ac{awgn} with entries having the variance $\sigma_{\rm RC}^2$. Similar to the derivation of \eqref{eq:received_all_pilot_output_new}, the signal received at the \ac{bs} in baseband representation at each $t$-th time slot for each $b$-th sub-frame can be expressed as follows:
\begin{equation}\label{eq:received_all_pilot_BS_new}
    \mathbf{y}_{\rm BS}\left( b,t \right)= \mathbf{H}_{\rm RB} { \boldsymbol  \Psi} \left( {\boldsymbol \rho}(b), {\boldsymbol \psi}(b)\right)  \mathbf{G} \mathbf{s}(t) +  \mathbf{z}_{\rm BS}\left( b,t \right),
\end{equation}
where $\mathbf{z}_{\rm BS}\left( b,t \right) \in\mathbb{C}^{M }$ is a zero-mean \ac{awgn} having i.i.d. elements each with variance $\sigma_{\rm BS}^2$.

Let $\mathbf{y}_{\rm RC}\left( b \right)\triangleq\left[\mathbf{y}_{\rm RC}\left( b,1 \right),\mathbf{y}_{\rm RC}\left( b,2 \right),\ldots,\mathbf{y}_{\rm RC}\left( b,T \right)\right] \in \mathbb{C}^{N_r \times T}$ be the matrix collecting the received signals at the HRIS over $T$ time slots for each $b$-th block, i.e.:
\begin{equation}\label{eq:y_rc}
\mathbf{y}_{\rm RC}\left( b \right) = {\boldsymbol \Phi}\left( {\boldsymbol \rho}(b), {\boldsymbol \phi}(b)\right) \mathbf{G} \mathbf{s} + \mathbf{z}_{\rm RC}\left( b \right),
\end{equation}
where $\mathbf{z}_{\rm RC}\left( b \right)\triangleq\left[\mathbf{z}_{\rm RC}\left( b,1 \right),\mathbf{z}_{\rm RC}\left( b,2 \right),\ldots,\mathbf{z}_{\rm RC}\left( b,T \right)\right] \in \mathbb{C}^{N_r \times T} $, and similarly define $\myMat{S}\triangleq\left[\mathbf{s}(1),\mathbf{s}(2),\ldots,\mathbf{s}(T) \right]$, for which it holds that $\myMat{S}\myMat{S}^{\rm H} = T{\bf I}_K$. Let also the definition of the $N_r\, B\times T$ matrix $\mathbf{y}_{\rm RC}$, generated by stacking the rows of the $B$ matrices $\mathbf{y}_{\rm RC}\left( 1 \right),\mathbf{y}_{\rm RC}\left( 2\right),\ldots, \mathbf{y}_{\rm RC}\left( B\right)$. From \eqref{eq:y_rc}, $\mathbf{y}_{\rm RC}$ can be written as a linear function of the HRIS-\acp{ut} channel $\mathbf{G}$ as follows:
\begin{equation}
    \label{eqn:LinearRC}
    \mathbf{y}_{\rm RC} = \mathbf{A}_{\rm RC}\left( \left\{\myVec{\rho}(b), \myVec{\phi}(b)\right\}\right) {\mathbf{G}} \mathbf{s} + \mathbf{z}_{\rm RC},
\end{equation}
where $\mathbf{z}_{\rm RC} \in \mathbb{C}^{N_r\, B\times T}$ is the row stacking of the matrices $\mathbf{z}_{\rm RC}(1),\mathbf{z}_{\rm RC}(2),\ldots, \mathbf{z}_{\rm RC}(B)$, while the matrix $\mathbf{A}_{\rm RC} \in \mathbb{C}^{N_r\, B\times N}$ is defined as:
\begin{equation}\label{eq:Arc}
   \mathbf{A}_{\rm RC}\left( \left\{\myVec{\rho}(b), \myVec{\phi}(b)\right\}\right) \triangleq \left[\myMat{\Phi}\left( \myVec{\rho}(1), \myVec{\phi}(1)\right)^{\rm T},\ldots,\myMat{\Phi}\left( \myVec{\rho}(B), \myVec{\phi}(B)\right)^{\rm T}\right]^{\rm T}. 
\end{equation}
Similarly, by letting $\mathbf{y}_{\rm BS}\left( b \right)\triangleq\left[\mathbf{y}_{\rm BS}\left( b,1 \right),\mathbf{y}_{\rm BS}\left( b,2 \right),\ldots,\mathbf{y}_{\rm BS}\left( b,T \right)\right] \in \mathbb{C}^{M \times T}$ be the matrix including  the received signals at BS during the $T$ time slots for each $b$-th sub-frame, the following expression is deduced:
\begin{equation}\label{eq:y_BS}
\mathbf{y}_{\rm BS}\left( b \right) = \mathbf{H}_{\rm RB} {\boldsymbol \Psi}\left( {\boldsymbol \rho}(b), {\boldsymbol \psi}(b)\right) \mathbf{G} \mathbf{s} + \mathbf{z}_{\rm BS}\left( b \right),
\end{equation}
where $\mathbf{z}_{\rm BS}\left( b \right)\triangleq\left[\mathbf{z}_{\rm BS}\left( b,1 \right),\mathbf{z}_{\rm BS}\left(b,2 \right),\ldots,\mathbf{z}_{\rm BS}\left( b,T \right)\right] \in \mathbb{C}^{M \times T} $.

As in conventional \ac{ris}-empowered communication systems, e.g., \cite{huang2019reconfigurable,RIS_challenges_all,10600711,wu2019intelligent}, it is herein assumed that the \ac{bs} maintains a high-throughput direct link with the HRIS. For conventional solely reflective and almost passive \acp{ris}, this link is used for controlling the \ac{ris} reflection pattern. In HRISs, which possess signal reception capability, this link is also used for conveying valuable information from the HRIS to the \ac{bs}. Therefore, inspired by the recent discussions for autonomous RISs with basic computing and storage capabilities \cite{10237986,HRIS_Mag_all,self_configuring_RIS,DRL_automonous_RIS,SORIS}, the focus later on will be on channel estimation carried out at both the HRIS side as well as the \ac{bs}. In particular, the goal is to characterize the achievable \ac{mse} in recovering the HRIS-\acp{ut} channel $\mathbf{G}$ from \eqref{eqn:LinearRC}, along with the \ac{mse} in estimating $\mathbf{H}_{\rm BR}$ at the \ac{bs} side from \eqref{eq:y_BS} and from the estimate of $\mathbf{G}$, denoted $\hat{\mathbf{G}}$, provided by the HRIS. The pilot matrix $\mathbf{s}$ in \eqref{eqn:LinearRC} and \eqref{eq:y_BS} is assumed to be known at both the HRIS and \ac{bs}. It is also noted that different HRIS configurations $\left\{\myVec{\rho}(b), \myVec{\phi}(b), \myVec{\psi}(b)\right\}$ result in different pilot signal strengths in \eqref{eqn:LinearRC} and \eqref{eq:y_BS}. Therefore, the focus will also be on the configuration of the HRIS controllable parameters in order to facilitate channel estimation based on the characterized \ac{mse}.

\subsubsection{Channel Estimation for Noise-Free Channels}
Let us first consider communications carried out in the case of absence of noise, i.e., the noise terms in \eqref{eq:received_all_pilot_output_new} and \eqref{eq:received_all_pilot_BS_new} are set to be zero, i.e., $\sigma_{\rm RC}^2=\sigma_{\rm BS}^2=0$. In such scenarios, one should be able to fully recover both $\mathbf{H}_{\rm BR}$ and $\mathbf{G}$ from the observed signals $\mathbf{y}_{\rm RC}(n)$ and $\mathbf{y}_{\rm BS}(n)$. The number of pilots required to achieve  accurate recovery is stated in the following proposition. 
\begin{proposition}
\label{pro:Noiseless}
In the case of noise absence, $\mathbf{H}_{\rm BR}$ and $\mathbf{G}$ can be accurately recovered when the number of pilots $\tau$ satisfies the following inequality:
\begin{equation}
    \label{eqn:Noiseless}
    \tau \geq N \max\left\{1,KN_r^{-1}\right\}.
\end{equation}
\end{proposition}
\begin{proof}
By discarding the noise term in \eqref{eqn:LinearRC}, the received pilot signal at the HRIS is given by the expresion:
\begin{equation}
    \label{eqn:LinearRC_w/o}
    \mathbf{y}_{\rm RC} = \mathbf{A}_{\rm RC}\left( \left\{\myVec{\rho}(b), \myVec{\phi}(b)\right\}\right) {\mathbf{G}} \mathbf{s}.
\end{equation}
By using the identity $\operatorname{vec}(\mathbf{A B C})=\left(\mathbf{C}^{\rm T} \otimes \mathbf{A}\right) \operatorname{vec}(\mathbf{B})$, \eqref{eqn:LinearRC_w/o} can be rewritten as follows:
\begin{equation}
    \label{eqn:LinearRC_without}
    \vecc\left(\mathbf{y}_{\rm RC}\right) =\left( \mathbf{s}^{\rm T} \otimes \mathbf{A}_{\rm RC}\left( \left\{\myVec{\rho}(b), \myVec{\phi}(b)\right\}\right) \right){\rm vec}(\mathbf{G}).
\end{equation}
In the latter expression, the matrix definition $\mathbf{A}_{\rm 1} \triangleq \mathbf{s}^{\rm T} \otimes \mathbf{A}_{\rm RC}\left( \left\{\myVec{\rho}(b), \myVec{\phi}(b)\right\}\right)$ has been used. If $\mathbf{A}_{\rm 1}$ is a full-column-rank matrix, then $\operatorname{vec}\left(\mathbf{G}\right)$ can be recovered from \eqref{eqn:LinearRC_without} as follows:
\begin{equation}\label{eq:estimation_G_vector}
\operatorname{vec}\left(\mathbf{G}\right) = \mathbf{A}_{\rm 1}^{\dagger} \vecc\left(\mathbf{y}_{\rm RC}\right),
\end{equation}
where $\mathbf{A}_{\rm 1}^{\dagger}=\left(\mathbf{A}_{\rm 1}^{\rm H} \mathbf{A}_{\rm 1}\right)^{-1}\mathbf{A}_{\rm 1}^{\rm H}$ is the pseudoinverse of $\mathbf{A}_{\rm 1}$. Once the perfect estimate of $\operatorname{vec}\left(\mathbf{G}\right)$ is obtained, then the channel matrix ${\bf G}$ can be recovered accordingly. The dimension of $\mathbf{A}_{\rm 1}$ is $N_r \tau$ by $N K$, and recall that $\tau=BT$. Thus, in order to guarantee that $\mathbf{A}_{\rm 1}$ has a full column rank, the pilot length $\tau$ should satisfy the following inequality:
\begin{equation}\label{eq:pilot_length_1}
\tau \geq  \frac{NK}{N_r}.
\end{equation} 

On the other hand, by discarding the noise term in \eqref{eq:y_BS}, the received pilot signal at the BS during $T$ time slots for each $b$-th sub-frame can be expressed as follows:
\begin{equation}\label{eq:y_BS_w/o}
\mathbf{y}_{\rm BS}\left( b \right) = \mathbf{H}_{\rm RB} {\boldsymbol \Psi}\left( {\boldsymbol \rho}(b), {\boldsymbol \psi}(b)\right) \mathbf{G} \mathbf{s},
\end{equation}
or equivalently,
\begin{equation}\label{eq:y_BS_w/o2}
\vecc\left(\mathbf{y}_{\rm BS}\left( b \right)\right) =\left( \left( {\boldsymbol \Psi}\left( {\boldsymbol \rho}(b), {\boldsymbol \psi}(b)\right) \mathbf{G} \mathbf{s}\right)^{\rm T} \otimes \mathbf{I}_M \right) {\rm vec}(\mathbf{H}_{\rm BR}).
\end{equation}
In the sequel, the notation $\myVec{\bar y}_{\rm BS}$ for the $M\tau\times 1$ vector generated by stacking the vectors $\vecc \left(\myVec{ y}_{\rm BS}\left( 1 \right)\right),\vecc \left(\mathbf{y}_{\rm BS}\left( 2\right)\right),\ldots, \vecc \left(\mathbf{y}_{\rm BS}\left( B \right)\right)$ is used. It follows from \eqref{eq:y_BS_w/o2} that one can express $\mathbf{y}_{\rm BS}$ as
    $\myVec{\bar y}_{\rm BS} = \mathbf{ A}_{\rm 2}\vecc \left(\mathbf{H}_{\rm BR}\right)$,
where $\mathbf{A}_{\rm 2} \in \mathbb{C}^{M\,\tau \times M\,N}$ is given by 
\begin{equation} \label{eq:Ab}
   \mathbf{A}_{\rm 2} = \left[ {\boldsymbol \Psi}\left( {\boldsymbol \rho}(1), {\boldsymbol \psi}(1)\right) \mathbf{G} \mathbf{s},\ldots, {\boldsymbol \Psi}\left( {\boldsymbol \rho}(B), {\boldsymbol \psi}(B)\right) \mathbf{G} \mathbf{s}\right]^{\rm T}\otimes \mathbf{I}_M.
\end{equation}
Similar to $\mathbf{A}_{\rm 1}$ before, ${\rm vec}(\mathbf{H}_{\rm BR})$ can be perfectly recovered if $\mathbf{A}_{\rm 2}$ is a full-column-rank matrix, i.e., it holds:
%
${\rm vec}(\mathbf{H}_{\rm BR}) =\mathbf{A}_{\rm 2}^{\dagger} \myVec{\bar y}_{\rm BS}$,
%
where $\mathbf{A}_{\rm 2}^{\dagger} =\left(\mathbf{A}_{\rm 2}^{\rm H} \mathbf{A}_{\rm 2}\right)^{-1}\mathbf{A}_{\rm 2}^{\rm H}$. In order to guarantee that $\mathbf{A}_{\rm 2}$ has a full column rank, the pilot length $\tau$ should satisfy the following inequality:
\begin{equation}\label{eq:pilot_length_2}
\tau \geq  N.
\end{equation}

By putting \eqref{eq:pilot_length_1} and \eqref{eq:pilot_length_2} together, it is concluded that the number of pilots $\tau$ should satisfy the inequality:
\begin{equation}
    \label{eqn:Pilot_length}
    \tau \geq N \max\left\{1,\frac{K}{N_r}\right\},
\end{equation}
which completes the proof of Proposition \ref{pro:Noiseless}. 
\end{proof} 

Proposition \ref{pro:Noiseless} demonstrates the intuitive benefit of HRISs in facilitating individual channel estimation with reduced number of pilots, as compared to existing techniques for estimating the cascaded \acp{ut}-HRIS-\ac{bs} channels (e.g., \cite{Tsinghua_RIS_Tutorial,wang2020channel}). For instance, for a multi-user \ac{mimo} system with $M=16$ antennas at the \ac{bs}, $N_r=8$ reception RF chains at the HRIS, $N=64$ HRIS elements, and $K=8$ \acp{ut}, the adoption of an HRIS allows recovering $\mathbf{H}_{\rm BR}$ and $\mathbf{G}$ separately using $\tau=64$ pilots. By contrast, the method proposed in \cite{wang2020channel} requires transmitting over $90$ pilots to identify the cascaded channel coefficients $[\mathbf{H}_{\rm BR}]_{m,l} [\mathbf{G}]_{l,k}$ $\forall l,k$ and $m=1,2,\ldots,M$. This reduction in pilot signals is directly translated into improved spectral efficiency, as less pilots are to be transmitted in each coherence duration.

\subsubsection{Channel Estimation for Noisy Channels}
The characterization of the number of required pilots in Proposition \ref{pro:Noiseless} provides an initial understanding of the HRIS's capability in providing efficient channel estimation. However, as Proposition \ref{pro:Noiseless} considers an effectively noise-free setup, it is invariant of the fact that HRISs split the power of their received signal $\mathbf{R}(n)$ between the reflected and received components. In the presence of noise, this division of the signal power may result in \ac{snr} degradation. Therefore, channel estimation using HRISs in the presence of noise is studied in the sequel, quantifying the achievable \ac{mse} in estimating the individual HRIS-\acp{ut} and \ac{bs}-HRIS channels for a fixed HRIS configuration $\left\{\myVec{\rho}(b), \myVec{\phi}(b), \myVec{\psi}(b)\right\}$. 
For notation brevity, in the following, the set of HRIS parameters affecting its reception is defined as $\myMat{\Phi} \triangleq \left\{\myVec{\rho}(b), \myVec{\phi}(b)\right\}$, the HRIS parameters affecting the reception at the \ac{bs} as $\myMat{\Psi}(b) \triangleq \myMat{\Psi}(\myVec{\rho}(b),\myVec{\psi}(b))$, and the overall HRIS parameters as $\myMat{\Omega} \triangleq \left\{ \myVec{\rho}(b), \myVec{\phi}(b), \myVec{\psi}(b)\right\}$. The assumption that the thermal noise powers at the HRIS and \ac{bs} reception units are of the same level, i.e., $\sigma_{\rm BS}^2 = \sigma_{\rm RC}^2=\sigma^2$, is also made.

In the following, the achievable \ac{mse} performance in recovering $\mathbf{G}$ at the HRIS, using its locally collected measurements for a given HRIS parameterization, denoted by $\mathcal{ E}_{\mathbf{G}} \left( \myMat{\Phi} \right)$, is characterized.

\begin{theorem}\label{Theorem:1}
The HRIS-\acp{ut} channel $\mathbf{G}$ can be recovered with the following \ac{mse} performance:
\begin{equation*}
\mathcal{E}_{\mathbf{G}} \left( \myMat{\Phi} \right)  = {\rm Tr} \left\{\left(\mathbf{R}_{\rm G}^{-1} +T\frac{ \Gamma}{K}\mathbf{A}_{\rm RC}\left( \myMat{\Phi} \right)^{\rm H} \mathbf{A}_{\rm RC}\left( \myMat{\Phi} \right)\right)^{-1}\right\}
\end{equation*}
where $\mathbf{R}_{\rm G} \triangleq \left(\sum_{k=1}^K g_k\right) \mathbf{I}_N$ and  $\Gamma\triangleq\frac{P_t}{\sigma^2}$ represents the transmit SNR with $P_t$ denoting each UT's power used for transmitting the pilot symbols.
\end{theorem}
\begin{proof}

For notation brevity, the definition $\myMat{\Phi} \triangleq \left\{\myVec{\rho}(b), \myVec{\phi}(b)\right\}$ is used during the proof of Theorem \ref{Theorem:1}. To estimate the channels between the HRIS and the UTs, $\mathbf{y}_{\rm RC}$ in \eqref{eqn:LinearRC} is projected on $\mathbf{s}^{\rm H}$, yielding the expression:
\begin{equation}\label{eq:estimate_g_k}
    \mathbf{\tilde y}_{\rm RC}=\frac{1}{T} \mathbf{y}_{\rm RC} \mathbf{s}^{\rm H}= \mathbf{A}_{\rm RC}\left( \myMat{\Phi}\right) \mathbf{G} + \mathbf{\tilde z}_{\rm RC},
\end{equation}
where $\mathbf{\tilde z}_{\rm RC} \triangleq \frac{1}{T}\mathbf{z}_{\rm RC}\mathbf{s}^{\rm H}$, whose distribution is given by $C \mathcal{N}\left(0, K(T \Gamma)^{-1} \myVec{ I}_{N_r B}\right) $. According to \eqref{eq:estimate_g_k}, the linear estimation that minimizes the MSE of the estimation of $\mathbf{G}$ has the following form \cite{biguesh2006training}:
\begin{equation}\label{eq:LMMSE_g_k}
   \mathbf{\hat G}= \mathbf{M}_{\rm opt} \mathbf{\tilde y}_{\rm RC},
\end{equation}
where $\mathbf{M}_{\rm opt}$ is the linear estimator, which can be obtained by solving the following problem:
\begin{equation}\label{eq:MMSE_estimator}
\begin{split}
\mathbf{M}_{\rm opt} & \triangleq\arg \min _{\mathbf{M}} {\mathbb{E}} \left\{\left\|\mathbf{G}-{\mathbf{\hat G}}\right\|_{F}^{2} \right\} = \arg \min _{{\mathbf{M}}} {\mathbb{E}} \left\{\left\|\mathbf{G} - \mathbf{M} \mathbf{\tilde y}_{\rm RC}\right\|_{F}^{2}\right\}.
\end{split}
\end{equation}
The error of this estimator can be given by 
\begin{equation}\label{eq:MSE}
\begin{split}
&\mathcal{ E}_{\mathbf{G}} \left(\myMat{\Phi}\right) = \mathbb{E} \left\{\left\|\mathbf{G} - \mathbf{M} \mathbf{\tilde y}_{\rm RC}\right\|_{F}^{2}\right\}  \\ 
&= {\rm Tr} \left(\mathbf{R}_{\rm G}\right) - {\rm Tr} \left(\mathbf{R}_{\rm G} \mathbf{A}_{\rm RC}\left( \myMat{\Phi}\right)^{\rm H} {\mathbf{M}}^{\rm H} \right) - {\rm Tr} \left({\mathbf{M}} \mathbf{A}_{\rm RC}\left( \myMat{\Phi}\right)\, \mathbf{R}_{\rm G}\right) \\
&+ {\rm Tr} \left({\mathbf{M}} \left( \mathbf{A}_{\rm RC}\left( \myMat{\Phi}\right) \mathbf{R}_{\rm G} \mathbf{A}_{\rm RC}\left( \myMat{\Phi}\right)^{\rm H} + K (T \Gamma)^{-1} {\myVec{ I}}_{N_r B} \right)\,{\mathbf{M}}^{\rm H} \right),
\end{split}
\end{equation}
where $ \mathbf{R}_{\rm G} = \mathbb{E}\left[{\mathbf{G}} {\mathbf{G}}^{\rm H} \right] = \left(\sum_{k=1}^{K} g_k\right) \myVec{ I}_N$ denotes the covariance matrix of ${\mathbf{G}}$. Since the second-order channel statistics vary slowly with time in general, it is here assumed that $ \mathbf{R}_{\rm G}$ can be perfectly estimated at the HRIS. The optimal $\mathbf{M}_{opt}$ can be found from $\partial \mathcal{ E}_{\mathbf{G} } \left(\myMat{\Phi}\right) / \partial \mathbf{M} = 0$
and is given by
\begin{equation} \label{eq:optimal_MSE_G}
\begin{split}
 \mathbf{M}_{\rm opt} =   \mathbf{R}_{\rm G}\,  \mathbf{A}_{\rm RC}\left( \myMat{\Phi}\right)^{\rm H} \bigg( \mathbf{A}_{\rm RC}\left( \myMat{\Phi}\right) \mathbf{R}_{\rm G} \mathbf{A}_{\rm RC}\left( \myMat{\Phi}\right)^{\rm H} +K (T \Gamma)^{-1}{\mathbf{I}_{N_r B}} \bigg)^{-1}.
 \end{split}
\end{equation}     
Substituting \eqref{eq:optimal_MSE_G} into \eqref{eq:MSE} and using the matrix inversion lemma, the Minimum MSE (MMSE) estimation error of $\mathbf{G}$ can be derived as follows:
\begin{equation}\label{eq:MMSE_error_G}
\mathcal{ E}_{\mathbf{G}} \left(\myMat{\Phi}\right)  = {\rm Tr} \left\{\left(\mathbf{R}_{\rm G}^{-1} +T \Gamma\, K^{-1}\mathbf{A}_{\rm RC}\left( \myMat{\Phi}\right)^{\rm H} \mathbf{A}_{\rm RC}\left( \myMat{\Phi}\right)\right)^{-1}\right\}.
\end{equation}

The linear MMSE estimator of $\mathbf{G}$ can be expressed as:
\begin{equation}\label{eq:LMMSE_G}
\begin{split}
   {\mathbf{\hat G}}=  \mathbf{R}_{\rm G}  \mathbf{A}_{\rm RC}\left( \myMat{\Phi}\right)^{\rm H} & \bigg( \mathbf{A}_{\rm RC}\left( \myMat{\Phi}\right) \mathbf{R}_{\rm G} \mathbf{A}_{\rm RC}\left( \myMat{\Phi}\right)^{\rm H} + K(T \Gamma)^{-1}{\myVec{ I}}_{N_r B} \bigg)^{-1}\mathbf{\tilde y}_{\rm RC},
\end{split}
\end{equation}
and it is easy to verify that the mean of ${\mathbf{\hat G}}$ is zero, i.e., $ \mathbb{E} \left\{ {\mathbf{\hat G}} \right\} =0$. Its covariance matrix is given by the following expression:
\begin{align}\label{eq:COvariance_G}
{\boldsymbol \Sigma} \left(\myMat{\Phi}\right) & = \mathbb{E} \left\{ {\mathbf{\hat G}} {\mathbf{\hat G}}^{\rm H} \right\}\\ 
& = \mathbf{R}_{\rm G}  \mathbf{A}_{\rm RC}\left( \myMat{\Phi}\right)^{\rm H} \bigg( \mathbf{A}_{\rm RC}\left( \myMat{\Phi}\right) \mathbf{R}_{\rm G} \mathbf{A}_{\rm RC}\left( \myMat{\Phi}\right)^{\rm H} +K  (T \Gamma)^{-1}{\myVec{ I}}_{N_r B} \bigg)^{-1} \mathbf{A}_{\rm RC}\left( \myMat{\Phi}\right) \mathbf{R}_{\rm G}^{\rm H}.\nonumber
\end{align}
Finally, by defining the channel estimation error as $\mathbf{\tilde G} \triangleq \mathbf{G} - \mathbf{\hat G}$ denote the channel estimation error, it can be easily verified that it has zero mean and the covariance matrix:
\begin{equation}\label{eq:error_MSE}
\mathbf{R}_{\bf \tilde G} \left(\myMat{\Phi}\right)  
= \left(\mathbf{R}_{\rm G}^{-1} + T \frac{\Gamma}{K}\mathbf{A}_{\rm RC}\left( \myMat{\Phi}\right)^{\rm H} \mathbf{A}_{\rm RC}\left( \myMat{\Phi}\right)\right)^{-1}.
\end{equation}
The latter expression concludes the proof.
\end{proof} 

Theorem~\ref{Theorem:1} allows to compute the achievable \ac{mse} in estimating $\mathbf{G}$ at the HRIS side for a given configuration of its reception phase profile, determined by $\myMat{\Phi} $, i.e., by $ \myVec{\rho}(b)$ and $\myVec{\phi}(b)$.

\begin{lemma}\label{lemma:1}
Let $\mathbf{\hat G}$ and $\mathbf{\tilde G} = \mathbf{G} - \mathbf{\hat G}$ denote the estimation of $\mathbf{G}$ and its corresponding estimation error, respectively. The distributions of $\mathbf{\hat G}$ and $\mathbf{\tilde G}$ are respectively given by
\begin{equation*}
\mathbf{\hat G}    \sim \mathcal{C} \mathcal{N} \left(0,{\boldsymbol \Sigma} \left(\myMat{\Phi}\right) \right), \qquad
\mathbf{\tilde G}    \sim \mathcal{C} \mathcal{N} \left(0,\mathbf{R}_{\bf \tilde G} \left(\myMat{\Phi}\right)\right),
\end{equation*}
where $\mathbf{R}_{\bf \tilde G} \left(\myMat{\Phi}\right) $ and ${\boldsymbol \Sigma} \left(\myMat{\Phi}\right) $ are defined as follows:
\begin{equation*}
\mathbf{R}_{\bf \tilde G} \left(\myMat{\Phi}\right)  \triangleq  
\left(\mathbf{R}_{\rm G}^{-1} + T \Gamma K^{-1}\mathbf{A}_{\rm RC}\left( \myMat{\Phi}\right)^{\rm H} \mathbf{A}_{\rm RC}\left( \myMat{\Phi}\right)\right)^{-1}, 
\end{equation*}
and 
\begin{equation*}
{\boldsymbol \Sigma} \triangleq \mathbf{R}_{\rm G}  \mathbf{A}_{\rm RC}\left( \myMat{\Phi}\right)^{\rm H}
 \Bigg( \mathbf{A}_{\rm RC}\left( \myMat{\Phi}\right) \mathbf{R}_{\rm G} \mathbf{A}_{\rm RC}\left(\myMat{\Phi}\right)^{\rm H} + K (T \Gamma)^{-1}{\mathbf{I}_{N_rB}} \Bigg)^{-1} \mathbf{A}_{\rm RC}\left( \myMat{\Phi}\right) \mathbf{R}_{\rm G}^{\rm H}. 
\end{equation*}
 \end{lemma}

\begin{theorem}\label{Theorem:2}
The \ac{bs}-HRIS channel $\mathbf{H}_{\rm BR}$ can be recovered with the following \ac{mse} performance:
\begin{align*}  
&\mathcal{E}_{\mathbf{H}_{\rm BR}}\left(\myMat{\Omega} \right) = {\rm Tr} \left(\left(\frac{1}{\beta} \myMat{I}_{M N} + \left(K \sum_{j=1}^B \sum_{i=1}^B{\rm Tr}\left(\left[\mathbf{D}(\myMat{\Omega})^{-T}\right]_{i,j}\right)\, \myMat{\Psi}(i){\boldsymbol \Sigma} \left(\myMat{\Phi}\right) \myMat{\Psi}^{\rm H} (j)\right)^{\rm T}
\otimes \mathbf{I}_M\right)^{-1}\right),
\end{align*}
where $\mathbf{D}(\myMat{\Omega})$ is a ${BK \times BK}$ matrix which can be partitioned into $B \times B$ blocks with each block being a $K \times K$ submatrix. The $i$-th row and $j$-th column block of $\mathbf{D}(\myMat{\Omega})$ is defined as follows:
\begin{equation*}
\begin{split}
\left[\mathbf{D}\left(\myMat{\Omega} \right)\right]_{i,j} = \left\{\begin{array}{ll}
\frac{\beta}{K} {\rm Tr}\left(\myMat{\Psi}(j)^{\rm H}\myMat{\Psi}(i)\mathbf{R}_{\bf \tilde G} \left(\myMat{\Phi}\right)\right)\mathbf{I}_K + (T \Gamma)^{-1} \mathbf{I}_{ K}, & {\rm if}~ i = j \\
\frac{\beta}{K} {\rm Tr}\left(\myMat{\Psi}(j)^{\rm H}\myMat{\Psi}(i)\mathbf{R}_{\bf \tilde G} \left(\myMat{\Phi}\right)\right)\mathbf{I}_K, & {\rm if}~ i \neq j
\end{array}\right..
\end{split}
\end{equation*}
\end{theorem}
\begin{proof}

For notation brevity, the definitions $\myMat{\Phi} \triangleq \left\{\myVec{\rho}(b), \myVec{\phi}(b)\right\}$,  $\myMat{\Omega} \triangleq \left\{ \myVec{\rho}(b), \myVec{\phi}(b), \myVec{\psi}(b)\right\}$, and $\myMat{\Psi}(b) \triangleq \myMat{\Psi}(\myVec{\rho}(b),\myVec{\psi}(b))$ are used during the proof of Theorem~\ref{Theorem:2}. By projecting $\mathbf{y}_{\rm BS}\left( b \right)$ defined in \eqref{eq:y_BS} on $\mathbf{s}^{\rm H}$ and scaling the resulting term by $1/T$, the following expression is obtained: 
\begin{equation}\label{eq:estimation_BS_project}
\mathbf{\tilde y}_{\rm BS}\left( b \right) = \frac{1}{T} \mathbf{y}_{\rm BS}[l]  \mathbf{s}^{\rm H}=  \mathbf{H}_{\rm RB} {\boldsymbol \Psi}\left(b\right) \mathbf{G}  + \mathbf{\tilde z}_{\rm BS}\left( b \right),
\end{equation}
where $\mathbf{\tilde z}_{\rm BS}\left( b \right)\triangleq\frac{1}{T} \mathbf{ z}_{\rm BS}\left( b \right) \mathbf{s}^{\rm H}$.
By applying the identity $\operatorname{vec}(\mathbf{A B C})=\left(\mathbf{C}^{\rm T} \otimes \mathbf{A}\right) \operatorname{vec}(\mathbf{B})$, $\mathbf{\tilde y}_{\rm BS}[l]$ in \eqref{eq:estimation_BS_project} can be rewritten in the following vector form:
\begin{align}\label{eq:vector_form_error}
& \vecc \left(\mathbf{\tilde y}_{\rm BS}\left( b \right)\right) = \left(\mathbf{G}^{\rm T} {\boldsymbol \Psi}\left(b\right)^{\rm T} \otimes \mathbf{I}_{M}\right) \vecc \left(\mathbf{H}_{\rm BR}\right) +  \vecc \left(\mathbf{\tilde z}_{\rm BS}\left( b \right)\right)\\
& = \left(\mathbf{\hat G}^{\rm T} {\boldsymbol \Psi}\left(b\right)^{\rm T} \otimes \mathbf{I}_M\right) \vecc \left(\mathbf{H}_{\rm BR}\right) + \left(\mathbf{\tilde G}^{\rm T} {\boldsymbol \Psi}\left( b\right)^{\rm T} \otimes \mathbf{I}_M\right) \vecc \left(\mathbf{H}_{\rm BR}\right)   +  \vecc \left(\mathbf{\tilde z}_{\rm BS}\left( b \right)\right).\nonumber
\end{align}

Let the notation $\mathbf{y}_{\rm BS}$ represent the $MKB\times 1$ vector generated by stacking the following vectors: $\vecc \left(\mathbf{\tilde y}_{\rm BS}\left( 1 \right)\right),\vecc \left(\mathbf{\tilde y}_{\rm BS}\left( 2\right)\right),\ldots, \vecc \left(\mathbf{\tilde y}_{\rm BS}\left( b \right)\right)$. The vector $\mathbf{y}_{\rm BS}$ from \eqref{eq:vector_form_error} can be expressed as follows:
\begin{equation}
    \label{eqn:LinearBS}
    \mathbf{y}_{\rm BS} = \mathbf{A}_{\rm BS} {\mathbf{h}}   +  \underbrace{\Delta \mathbf{A}_{\rm BS} {\mathbf{h}} + \mathbf{z}_{\rm BS} }_{\triangleq\mathbf{z}},
\end{equation}
where $\mathbf{z}_{\rm BS}$ results from stacking $\vecc \left(\mathbf{\tilde z}_{\rm BS}\left( 1 \right)\right), \vecc \left(\mathbf{\tilde z}_{\rm BS}\left( 2\right)\right),\ldots,\vecc \left(\mathbf{\tilde z}_{\rm BS}\left( B \right)\right)$ and $\mathbf{h}\triangleq\vecc \left(\mathbf{H}_{\rm BR}\right)$, as well as $\mathbf{A}_{\rm BS} \in \mathbb{C}^{M\,K\, B\times M\,N}$ and $\Delta \mathbf{A}_{\rm BS} \in \mathbb{C}^{M\,K\, B\times M\,N}$ are respectively given by 
\begin{equation} \label{eq:Ab2}
   \mathbf{A}_{\rm BS} \triangleq \left[\myMat{\Psi}(1) \mathbf{\hat G},\cdots,\myMat{\Psi}(B) \mathbf{\hat G}\right]^{\rm T}\otimes \mathbf{I}_M,
\end{equation}
\begin{equation} \label{eq:Delta_Ab}
  \Delta \mathbf{A}_{\rm BS} \triangleq \left[\myMat{\Psi}(1) \mathbf{\tilde G},\cdots,\myMat{\Psi}(B)  \mathbf{\tilde G}\right]^{\rm T}\otimes \mathbf{I}_M.
\end{equation}

Let us next define $\mathbf{z} \triangleq \Delta \mathbf{A}_{\rm BS} {\mathbf{h}} + \mathbf{z}_{\rm BS}$ as the effective noise vector at the BS, which includes the colored interference forwarded from the HRIS and the local AWGN vector  $\mathbf{z}_{\rm BS}$. The following Proposition provides the second-order statistics of $\mathbf{z}$.
\begin{proposition} 
\label{Prop:noise_covariance}
The covariance matrix of $\mathbf{z}$ is given by $\mathbf{R_z}\left(\myMat{\Omega} \right) \triangleq \mathbb{E}\left\{\mathbf{z} \mathbf{z}^{\rm H}\right\} = {\bf D}\left(\myMat{\Omega} \right) \otimes {\bf I}_M$,
where $\mathbf{D}\left(\myMat{\Omega} \right)$ is a ${BK \times BK}$ matrix, which can be partitioned into $B \times B$ blocks with each block being a $K \times K$ submatrix. The $i$-th row and $j$-th column block of $\mathbf{D}\left(\myMat{\Omega} \right)$ is defined as follows:
\begin{equation}\label{eq:D_Omega}
\left[\mathbf{D}\left(\myMat{\Omega} \right)\right]_{i,j}  \!=\! \left\{\begin{array}{ll}
\frac{\beta}{K} {\rm Tr}\left(\myMat{\Psi}(j)^{\rm H}\myMat{\Psi}(i)\mathbf{R}_{\bf \tilde G} \left(\myMat{\Phi}\right)\right)\mathbf{I}_K + (T \Gamma)^{-1} \mathbf{I}_{ K}, & {\rm if}~ i = j \\
\frac{\beta}{K} {\rm Tr}\left(\myMat{\Psi}(j)^{\rm H}\myMat{\Psi}(i)\mathbf{R}_{\bf \tilde G} \left(\myMat{\Phi}\right)\right)\mathbf{I}_K, & {\rm if}~ i \neq j
\end{array}\right.. 
\end{equation}
\end{proposition}

For notation brevity, in the following, the simplified notations $\mathbf{R_z}$ and $\mathbf{D}$ are used to represent $\mathbf{R_z}\left(\myMat{\Omega} \right)$ and $\mathbf{D}\left(\myMat{\Omega} \right)$, respectively. In addition, the linear MMSE estimator to estimate $\mathbf{h}$ from $\mathbf{y}_{\rm BS}$ as $\mathbf{\hat h} = \mathbf{T} \mathbf{y}_{\rm BS}$ is used, where matrix $\mathbf{T}$ provides the optimal solution that minimizes the following channel estimation error:
\begin{equation}\label{eq:Vec_MSE}
\begin{split}
\mathcal{E}_{\mathbf{h}}= \mathbb{E} \left\{\left\|\mathbf{h} -\mathbf{\hat h}  \right\|^{2}\right\}  = \mathbb{E} \left\{\left\|\mathbf{h} -\mathbf{T} \mathbf{y}_{\rm BS} \right\|^{2}\right\}.
\end{split}
\end{equation}
It is well-known \cite{sengijpta1995fundamentals} that  the optimal $\mathbf{T}$ can be obtained as follows
\begin{equation}\label{eq:Vec_MSE_T}
\begin{split}
\mathbf{T} &=\mathbb{E}\left[\mathbf{h} \mathbf{y}_{\rm BS}^{\rm H}\right]\left(\mathbb{E}\left[\mathbf{y}_{\rm BS} \mathbf{y}_{\rm BS}^{\rm H}\right]\right)^{-1} = \mathbf{R_h}\, \mathbf{A}_{\rm BS}^{\rm H}\, \left(\mathbf{A}_{\rm BS}\, \mathbf{R_h}\, \mathbf{A}_{\rm BS}^{\rm H}\, + \mathbf{R_z} \right)^{-1},
\end{split}
\end{equation}
where $\mathbf{R_h} \triangleq \mathbb{E}\left\{\mathbf{h}\, \mathbf{h}^{\rm H}\right\} = \beta \mathbf{I}_{MN}$. By substituting \eqref{eq:Vec_MSE_T} into \eqref{eq:Vec_MSE}, the following expression for $\mathcal{E}_{\myVec{h}}$ is obtained:
\begin{align}
\mathcal{E}_{\mathbf{h}} &= \mathbb{E}_{\mathbf{h},\mathbf{\hat G}} \left\{\left\|\mathbf{h} - \mathbf{R_h}\, \mathbf{A}_{\rm BS}^{\rm H}\, \left(\mathbf{A}_{\rm BS}\, \mathbf{R_h}\, \mathbf{A}_{\rm BS}^{\rm H}\, + \mathbf{R_z} \right)^{-1} \mathbf{y}_{\rm BS} \right\|^{2}\right\} \notag\\ 
 & = \mathbb{E}_{\mathbf{\hat G}}\left\{{\rm Tr} \left(\left(\mathbf{R_h}^{-1} + \mathbf{A}_{\rm BS}^{\rm H}\, \mathbf{R_z}^{-1}\, \mathbf{A}_{\rm BS}\right)^{-1}\right)\right\} \label{eq:Vec_MSE_G}\\ 
&= \mathbb{E}_{\mathbf{\hat G}}\left\{{\rm Tr} \left(\left(\mathbf{R_h}^{-1} + \left( \sum_{j=1}^B \sum_{i=1}^B\,\myMat{\Psi}(i)\mathbf{\hat G} \left[\mathbf{D}^{-T}\right]_{i,j} \mathbf{\hat G}^{\rm H} \myMat{\Psi}(j)^{\rm H} \right)^{\rm T}
\otimes \mathbf{I}_M\right)^{-1}\right)\right\} \notag.
\end{align} 

Moreover, the following expression holds:
\begin{align}
\mathbb{E}_{\mathbf{\hat G}}\left\{\mathbf{\hat G} \left[\mathbf{D}^{-T}\right]_{i,j} \mathbf{\hat G}^{\rm H}\right\} 
&=\mathbb{E}_{\mathbf{N}}\left\{{\boldsymbol \Sigma} \left(\myMat{\Phi}\right)^{1/2}\mathbf{N} \left[\mathbf{D}^{-T}\right]_{i,j} \mathbf{N}^{\rm H} {\boldsymbol \Sigma} \left(\myMat{\Phi}\right)^{1/2}\right\}\notag \\
&=K\, {\rm Tr}\left(\left[\mathbf{D}^{-T}\right]_{i,j}\right) {\boldsymbol \Sigma} \left(\myMat{\Phi}\right),
\label{eq:temp1}
\end{align}
where ${\mathbf{N}} \in \mathbb{C}^{N \times K}$ denotes a random matrix distributed as $\mathbf{N} \sim \mathcal{C} \mathcal{N}\left(\mathbf{0},  \mathbf{I}_{N}\right)$. 

In order to efficiently design the reflection and reception weights at the HRIS, the MSE defined in \eqref{eq:Vec_MSE_G} is next approximated with a deterministic MSE, using a standard bounding technique. In particular, the following lower-bound for $\mathcal{E}_{\mathbf{h}}$ is defined:
\begin{align}
\mathcal{E}_{\mathbf{h}}  
& \overset{(a)}{\geq}   {\rm Tr} \left(\left(\mathbf{R_h}^{-1} + \left( \sum_{j=1}^B \sum_{i=1}^B\,\myMat{\Psi}(i)\mathbb{E}_{\mathbf{\hat G}}\left\{\mathbf{\hat G} \left[\mathbf{D}^{-T}\right]_{i,j} \mathbf{\hat G}^{\rm H} \right\} \myMat{\Psi}(j)^{\rm H} \right)^{\rm T}
\otimes \mathbf{I}_M\right)^{-1}\right) \notag\\
& \overset{(b)}{=}   {\rm Tr} \left(\left(\mathbf{R_h}^{-1} + \left(K \sum_{j=1}^B \sum_{i=1}^B\,{\rm Tr}\left(\left[\mathbf{D}^{-T}\right]_{i,j}\right)\, \myMat{\Psi}(i){\boldsymbol \Sigma} \left(\myMat{\Phi}\right) \myMat{\Psi}(j)^{\rm H} \right)^{\rm T}
\otimes \mathbf{I}_M\right)^{-1}\right), \label{eq:C11}
\end{align} 
where $(a)$ holds from the Jensen’s inequality and the fact that $\rm{Tr}\left(\mathbf{X}^{-1}\right)$ is a convex function with respect to $\mathbf{X}$. In addition, $(b)$ comes from \eqref{eq:temp1}. 

Putting all the above together, the lower bound of the channel estimation error at the BS is obtained as stated in Theorem~\ref{Theorem:2}, which concludes the proof. 
\end{proof}  

Theorems~\ref{Theorem:1} and~\ref{Theorem:2} allow to evaluate the achievable \ac{mse} for the recovery of the individual channels $\mathbf{G}$ and $\mathbf{H}_{\rm BR}$. The fact that these \acp{mse} are given as functions of the HRIS parameters $\myMat{\Omega}$ enables to numerically optimize the HRIS configuration for the estimation of the individual channels. Lemma \ref{lemma:1} provides the statistical results for the estimation of $\mathbf{G}$ with respect to the HRIS reconfigurable parameters $\myMat{\Phi}$, more specifically, to $\myVec{\rho}(b)$ and $\myVec{\phi}(b)$. This estimation can be used from the BS to estimate the channel matrix $\mathbf{H}_{\rm BR}$. In particular, letting the HRIS convey the estimation of $\mathbf{G}$ to the \ac{bs} (via their control link~\cite{RIS_challenges_all,10600711,RISE6G_COMMAG}) allows achieving the \ac{mse} in recovering $\mathbf{H}_{\rm BR}$, as described by means of the previous theorem. Recall that the BS observes the reflected portion of the signal at the output of the \ac{bs}-HRIS channel.
 
In the following, the \ac{mse} performance evaluation results reveal the fundamental trade-off between the ability to recover $\mathbf{G}$ and $\mathbf{H}_{\rm BR}$, which is dictated mostly by the parameter $\myVec{\rho}$ determining the portion of the impinging signal being reflected; the remaining portion is sensed and used for channel estimation at the HRIS side.

\subsubsection{Numerical Results}
As previously discussed, for the considered uplink HRIS-empowered wireless MIMO communication system with $K$ UTs in Fig.~\ref{fig:SystemModel2}, the H\ac{ris} can estimate the unknown $NK$ channel coefficients of which the  \acp{ut}-\ac{ris} channel $\mathbf{G}$ is comprised. Recall that this channel can be measured via spatial sampling through the HRIS anlod combining weights, using the $N_r$ digital signal reception paths. At the absence of noise, the H\ac{ris} is able to recover the matrix $\mathbf{G}$ from $NK / N_r$ pilots, while the \ac{bs} requires at least $N$ pilots to recover the BS-H\ac{ris} channel $\mathbf{H}_{\rm BR}$ from its observations and the estimate provided by the HRIS via the control link. For instance, when $K=8$ \acp{ut} communicate via an H\ac{ris} comprised $N=64$ meta-atoms, all attached to $N_r=8$ RF chains, both individual channels can be recovered from merely $64$ pilots. For comparison, state-of-the-art methods for conventional \acp{ris} would require over $90$ pilot transmissions for recovering solely the cascaded channel for the case of a $16$-antenna \ac{bs} with fully digital combining~\cite{wang2020channel}. This simplistic strategy for channel estimation at the HRIS side is beneficial, not only in noise-free channels, but also in noisy setups. However, one needs to also account for the fact that the power splitting between the reflected and sensed waveforms, carried out by the H\ac{ris}, affects the resulting \acp{snr} both at the metasurface and the \ac{bs}. In particular, in noisy setups where the H\ac{ris} estimates the composite HRIS-\acp{ut} channel locally and forwards it to the \ac{bs}, there is an inherent trade-off between the accuracy in estimating each of the individual channels, which is dictated by how the H\ac{ris} is configured to split the power of the incident wave between its reflected and absorbed portions.
\begin{figure}[!t]
    \centering
    \includegraphics[width=\columnwidth]{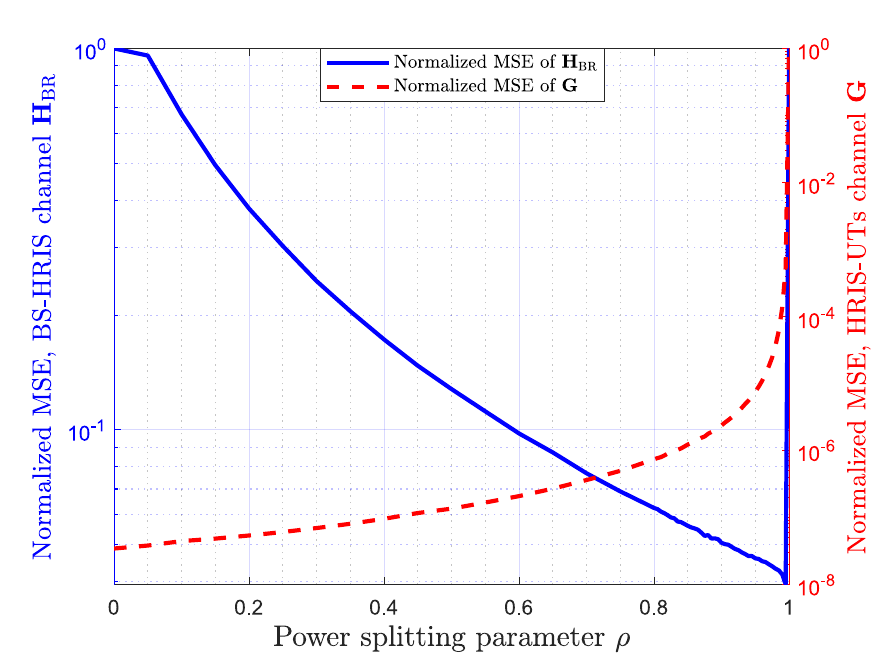}
    \caption{\small{Normalized \ac{mse} performance in recovering the combined HRIS-\acp{ut} channel $\mathbf{G}$ at the HRIS side and the \ac{bs}-HRIS channel $\mathbf{H}_{\rm BR}$ at the \ac{bs}, considering $30$~dB transmit SNR as well as different power splitting values $\rho$ and phase configurations.}}
    \label{fig:tradeoff2}
\end{figure}

In Fig.~\ref{fig:tradeoff2}, the aforedescribed trade-off between the \ac{mse} performances when estimating the HRIS-\acp{ut} channel at the H\ac{ris} and the BS-H\ac{ris} channel at the \ac{bs} is illustrated for the considered system ($M=16$, $K=8$, $N=64$, and $N_r=8$) for different values of the common power splitting parameter $\rho$ for all hybrid meta-atoms. Each curve corresponds to a different random setting of the individual phase shift at each meta-atom. The \acp{ut} have been uniformly distributed in a cell of $10$ meter radius, where the H\ac{ris} is located at the top edge of the cell, and at distance $50$ meters from the \ac{bs}. The \acp{ut} transmitted $T=70$ pilot symbols for channel estimation, which were received at the \ac{bs} after being reflected by the H\ac{ris} with \ac{snr} at $30$ dB. As it can be observed in the figure, there is a clear trade-off between the accuracy in estimating each of the individual channels, which is dictated by how the H\ac{ris} splits the power of the impinging signal. While the \ac{mse} values depend on the HRIS phase configuration, it is shown that, increasing the portion of the signal that is reflected in the range of up to $50\%$, notably improves the ability to estimate the BS-H\ac{ris} channel, while having only a minor effect on the \ac{mse} in estimating the H\ac{ris}-\acp{ut} channel. However, further increasing the amount of power reflected, notably degrades the \ac{mse} in estimating the HRIS-\acp{ut} channel, while hardly improving the accuracy of the BS-H\ac{ris} channel estimation.

\begin{figure}[!t]
    \centering
    \includegraphics[width=\columnwidth]{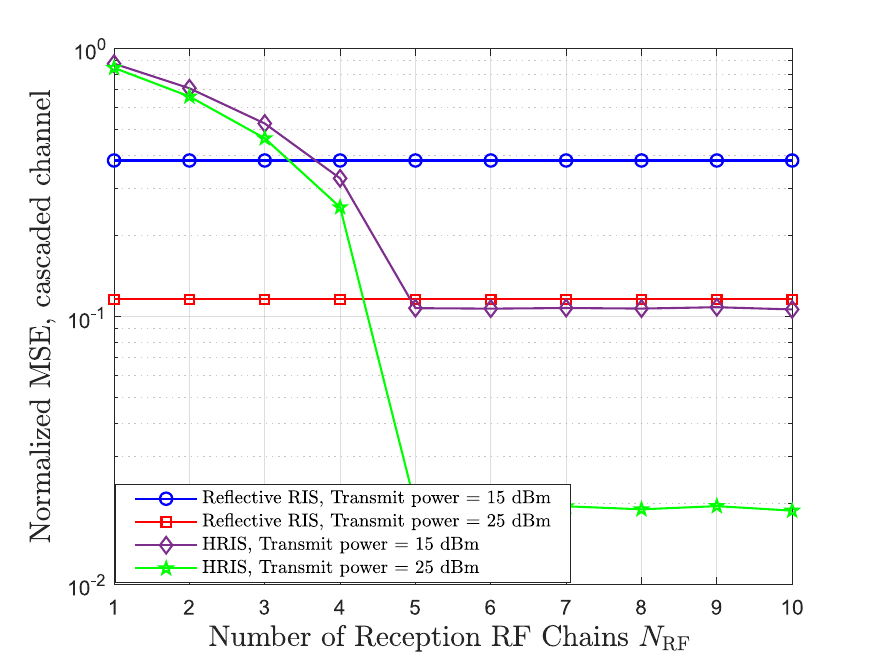}
    \caption{\small{Normalized \ac{mse} performance of the estimation od the cascaded channel as a function of the number $N_r$ of the reception RF chains at the HRIS for two different transmit SNR values in dB. The performance using a purely reflective RIS via the scheme of \cite{wang2020channel} is also shown.}}
    \label{fig:comparison}
\end{figure}

The ability to sense the individual channels at the HRIS side facilitates channel estimation, especially in wideband multi-user multi-antenna systems, where the conventional estimation of the cascaded BS-HRIS-UTs channel requires prohibitive overhead~\cite{wang2020channel}. In addition, it enables RF sensing, localization, and radio mapping whose role is envisioned to be prominent in 6G  \cite{Samsung}. 
Moreover, it can be beneficial even when one is interested in the cascaded channel. To show this, in Fig. \ref{fig:comparison}, the \ac{mse} performance in estimating the cascaded channel for the setup in Fig.~\ref{fig:tradeoff2} with an H\ac{ris}, which reflects on average {$50\%$} of the incoming signal, is depicted with respect to the number of reception RF chains $N_r$. The MSE values with the method presented in \cite{wang2020channel} for reflective \acp{ris} is also included for comparison.
As shown, the sensing capability of  H\acp{ris}  translates into improved  cascaded channel estimation, depending on how many RF chains it possesses,  though at the cost of higher power consumption and hardware complexity at the HRIS. 

\section{Conclusions}
\label{sec:Conclusions}
In this chapter, we reviewed the emerging concept of HRISs. In contrast to purely reflective RISs, HRISs can simultaneously reflect a portion of the impinging signal in a controllable manner, while enabling reception via tunable spatial sampling of the other portion of it. By presenting a simplified model for the HRIS dual functionality, and the respective reconfigurabilities, two indicative applications were discussed, one for simultaneous communications and sensing, and another that indicated their usefulness for estimating the individual channels in the uplink of a multi-user HRIS-empowered MIMO communication system. Representative numerical investigations showcased the HRIS potential for ISAC as well as channel estimation, with the latter application being particularly relevant to the HRIS self reconfigurability, and demonstrated the role of the power splitting parameter on the achievable performance. In addition, the interplay between the performance of the sensing/estimation and communications operations and the number of the HRIS reception RF chains was exhibited, with the latter parameter determining the overall power consumption and hardware complexity of the metasurface.

Apart from the capability of HRISs to significantly facilitate the integration and orchestration of metasurfaces in wireless networks, via their self-configuration potential~\cite{HRIS_Mag_all} which minimizes the dependence on external configuration and control, this new concept is lately gaining attention for localization~\cite{9726785,10124713,10557771}, sensing~\cite{HRIS_ISAC}, and RF mapping \cite{wymeersch2020radio,RFimaging}. This trend is a strong indication of the HRIS potential for ISAC, enabling simultaneous communications and radar target sensing~\cite{HRIS_ISAC_radarconf2025_1,HRIS_ISAC_radarconf2025_2}.

\bibliographystyle{IEEEtran}
\bibliography{IEEEabrv, refs}

\begin{thebibliography}{10}
\providecommand{\url}[1]{#1}
\csname url@samestyle\endcsname
\providecommand{\newblock}{\relax}
\providecommand{\bibinfo}[2]{#2}
\providecommand{\BIBentrySTDinterwordspacing}{\spaceskip=0pt\relax}
\providecommand{\BIBentryALTinterwordstretchfactor}{4}
\providecommand{\BIBentryALTinterwordspacing}{\spaceskip=\fontdimen2\font plus
\BIBentryALTinterwordstretchfactor\fontdimen3\font minus \fontdimen4\font\relax}
\providecommand{\BIBforeignlanguage}[2]{{%
\expandafter\ifx\csname l@#1\endcsname\relax
\typeout{** WARNING: IEEEtran.bst: No hyphenation pattern has been}%
\typeout{** loaded for the language `#1'. Using the pattern for}%
\typeout{** the default language instead.}%
\else
\language=\csname l@#1\endcsname
\fi
#2}}
\providecommand{\BIBdecl}{\relax}
\BIBdecl

\bibitem{Samsung}
``The next hyper- connected experience for all,'' White Paper, Samsung 6G Vision, Jun. 2020.

\bibitem{huang2019reconfigurable}
C.~Huang, A.~Zappone, G.~C. Alexandropoulos, M.~Debbah, and C.~Yuen, ``Reconfigurable intelligent surfaces for energy efficiency in wireless communication,'' \emph{IEEE Trans. Wireless Commun.}, vol.~18, no.~8, pp. 4157--4170, Aug. 2019.

\bibitem{di2019smart}
M.~Di~Renzo, M.~D. D.-T. Phan-Huy, A.~Zappone, M.-S. Alouini, C.~Yuen, V.~Sciancalepore, G.~C.~A. ans J.~Hoydis, H.~Gacanin, J.~de~Rosny, A.~Bounceu, G.~Lerosey, and M.~Fink, ``Smart radio environments empowered by reconfigurable {AI} meta-surfaces: an idea whose time has come,'' \emph{EURASIP J. Wireless Commun. Net.}, vol. 2019, no.~1, pp. 1--20, May 2019.

\bibitem{ETSI_RIS_COMSTD}
R.~Liu, S.~Zheng, Y.~Jiang, Q.~Wu, N.~Zhang, Y.~Liu, M.~Di~Renzo, and G.~C. Alexandropoulos, ``Sustainable wireless networks via reconfigurable intelligent surfaces ({RISs}): Overview of the {ETSI ISG RIS},'' \emph{IEEE Commun. Standards Mag.}, early access, 2025.

\bibitem{wu2021intelligent}
Q.~Wu, S.~Zhang, B.~Zheng, C.~You, and R.~Zhang, ``Intelligent reflecting surface-aided wireless communications: {A} tutorial,'' \emph{{IEEE} Trans. Commun.}, vol.~69, no.~5, pp. 3313--3351, 2021.

\bibitem{9673796}
V.~Jamali, G.~C. Alexandropoulos, R.~Schober, and H.~V. Poor, ``Low-to-zero-overhead {IRS} reconfiguration: Decoupling illumination and channel estimation,'' \emph{IEEE Commun. Lett.}, vol.~26, no.~4, pp. 932--936, 2022.

\bibitem{10670007}
K.~D. Katsanos, P.~D. Lorenzo, and G.~C. Alexandropoulos, ``Multi-{RIS}-empowered multiple access: A distributed sum-rate maximization approach,'' \emph{IEEE J. Sel. Topics Signal Process.}, vol.~18, no.~7, pp. 1324--1338, 2024.

\bibitem{9693982}
X.~Cao, B.~Yang, C.~Huang, G.~C. Alexandropoulos, C.~Yuen, Z.~Han, H.~V. Poor, and L.~Hanzo, ``Massive access of static and mobile users via reconfigurable intelligent surfaces: Protocol design and performance analysis,'' \emph{IEEE J. Sel. Areas Commun.}, vol.~40, no.~4, pp. 1253--1269, 2022.

\bibitem{Samarakoon_2020_all}
G.~C. Alexandropoulos, S.~Samarakoon, M.~Bennis, and M.~Debbah, ``Phase configuration learning in wireless networks with multiple reconfigurable intelligent surfaces,'' in \emph{Proc. IEEE GLOBECOM}, Taipei, Taiwan, Dec. 2020.

\bibitem{9406837}
D.~Selimis, K.~P. Peppas, G.~C. Alexandropoulos, and F.~I. Lazarakis, ``On the performance analysis of {RIS}-empowered communications over {N}akagami-$m$ fading,'' \emph{IEEE Commun. Lett.}, vol.~25, no.~7, pp. 2191--2195, 2021.

\bibitem{PLS_Kostas_all}
G.~C. Alexandropoulos, K.~Katsanos, M.~Wen, and D.~B. da~Costa, ``Safeguarding {MIMO} communications with reconfigurable metasurfaces and artificial noise,'' in \emph{Proc. IEEE ICC}, Montreal, Canada, Jun. 2021.

\bibitem{Counteracting}
G.~C. Alexandropoulos, K.~D. Katsanos, M.~Wen, and D.~B. Da~Costa, ``Counteracting eavesdropper attacks through reconfigurable intelligent surfaces: A new threat model and secrecy rate optimization,'' \emph{IEEE Open J. Commun. Society}, vol.~4, pp. 1285--1302, 2023.

\bibitem{RIS_challenges_all}
G.~C. Alexandropoulos, D.-T. Phan-Huy, K.~D. Katsanos, M.~Crozzoli, H.~Wymeersch, P.~Popovski, P.~Ratajczak, Y.~B{\'e}n{\'e}dic, M.-H. Hamon, S.~H. Gonzalez \emph{et~al.}, ``{RIS}-enabled smart wireless environments: Deployment scenarios, network architecture, bandwidth and area of influence,'' \emph{EURASIP J. Wireless Commun. and Netw.}, vol. 2023, no.~1, pp. 1--38, Oct. 2023.

\bibitem{10930892}
G.~Stamatelis, K.~Stylianopoulos, and G.~C. Alexandropoulos, ``Evolving multi-branch attention convolutional neural networks for online {RIS} configuration,'' \emph{IEEE Trans. Cogn. Commun. Netw.}, early access, 2025.

\bibitem{Tsinghua_RIS_Tutorial}
M.~Jian, G.~C. Alexandropoulos, E.~Basar, C.~Huang, R.~Liu, Y.~Liu, and C.~Yuen, ``Reconfigurable intelligent surfaces for wireless communications: {O}verview of hardware designs, channel models, and estimation techniques,'' \emph{Intell. Converged Netw.}, vol.~3, no.~1, pp. 1--32, Mar. 2022.

\bibitem{9758764}
R.~A. Tasci, F.~Kilinc, E.~Basar, and G.~C. Alexandropoulos, ``A new {RIS} architecture with a single power amplifier: Energy efficiency and error performance analysis,'' \emph{IEEE Access}, vol.~10, pp. 44\,804--44\,815, 2022.

\bibitem{9377648}
R.~Long, Y.-C. Liang, Y.~Pei, and E.~G. Larsson, ``Active reconfigurable intelligent surface-aided wireless communications,'' \emph{IEEE Trans. Wireless Commun.}, vol.~20, no.~8, pp. 4962--4975, 2021.

\bibitem{10596064}
E.~Basar, G.~C. Alexandropoulos, Y.~Liu, Q.~Wu, S.~Jin, C.~Yuen, O.~A. Dobre, and R.~Schober, ``Reconfigurable intelligent surfaces for 6g: Emerging hardware architectures, applications, and open challenges,'' \emph{IEEE Veh. Technol. Mag.}, vol.~19, no.~3, pp. 27--47, 2024.

\bibitem{10600711}
F.~Saggese, V.~Croisfelt, R.~Kotaba, K.~Stylianopoulos, G.~C. Alexandropoulos, and P.~Popovski, ``On the impact of control signaling in {RIS}-empowered wireless communications,'' \emph{IEEE Open J. Commun.Society}, vol.~5, pp. 4383--4399, 2024.

\bibitem{9827873}
G.~C. Alexandropoulos, V.~Jamali, R.~Schober, and H.~V. Poor, ``Near-field hierarchical beam management for {RIS}-enabled millimeter wave multi-antenna systems,'' in \emph{Proc. IEEE Sensor Array Multichannel Signal Process. Workshop}, 2022, pp. 460--464.

\bibitem{RISE6G_COMMAG}
E.~Calvanese~Strinati, G.~C. Alexandropoulos, H.~Wymeersch, B.~Denis, V.~Sciancalepore, R.~D'Errico, A.~Clemente, D.-T. Phan-Huy, E.~De~Carvalho, and P.~Popovski, ``Reconfigurable, intelligent, and sustainable wireless environments for {6G} smart connectivity,'' \emph{{IEEE} Commun. Mag.}, vol.~59, no.~10, pp. 99--105, Oct. 2021.

\bibitem{10802983}
F.~Saggese, V.~Croisfelt, K.~Stylianopoulos, G.~C. Alexandropoulos, and P.~Popovski, ``Control plane for reconfigurable intelligent surfaces,'' \emph{IEEE Commun. Standards Mag.}, vol.~8, no.~4, pp. 24--30, 2024.

\bibitem{wymeersch2020radio}
H.~Wymeersch \emph{et~al.}, ``Radio localization and mapping with reconfigurable intelligent surfaces: Challenges, opportunities, and research directions,'' \emph{{IEEE} Veh. Technol. Mag.}, vol.~15, no.~4, pp. 52--61, Dec. 2020.

\bibitem{RIS_Localization}
K.~Keykhosravi, B.~Denis, G.~C. Alexandropoulos, Z.~S. He, A.~Albanese, V.~Sciancalepore, and H.~Wymeersch, ``Leveraging {RIS}-enabled smart signal propagation for solving infeasible localization problems,'' \emph{IEEE Veh. Technol. Mag.}, vol.~18, no.~2, pp. 20--28, Jun. 2023.

\bibitem{taha2021enabling}
A.~Taha, M.~Alrabeiah, and A.~Alkhateeb, ``Enabling large intelligent surfaces with compressive sensing and deep learning,'' \emph{{IEEE} Access}, vol.~9, pp. 44\,304--44\,321, 2021.

\bibitem{HRIS_design_eucap2025}
M.~Birari, D.~S. Nagarkoti, A.~Katsounaros, H.~K.~R. Mudireddy, J.~Malik, and G.~C. Alexandropoulos, ``Dual-dielectric metasurface for simultaneous sensing and reconfigurable reflections,'' in \emph{Proc. European Conf. Antennas Propag.}, Stockholm, Sweden, 2025.

\bibitem{shlezinger2020dynamic}
N.~Shlezinger \emph{et~al.}, ``Dynamic metasurface antennas for {6G} extreme massive {MIMO} communications,'' \emph{{IEEE} Wireless Commun.}, vol.~28, no.~2, pp. 106--113, Apr. 2021.

\bibitem{9847609}
L.~You, J.~Xu, G.~C. Alexandropoulos, J.~Wang, W.~Wang, and X.~Gao, ``Energy efficiency maximization of massive {MIMO} communications with dynamic metasurface antennas,'' \emph{IEEE Trans. Wireless Commun.}, vol.~22, no.~1, pp. 393--407, 2023.

\bibitem{Gong_HMIMO_2023}
T.~Gong, P.~Gavriilidis, R.~Ji, C.~Huang, G.~C. Alexandropoulos, L.~Wei, M.~Debbah, H.~V. Poor, and C.~Yuen, ``Holographic {MIMO} communications: {T}heoretical foundations, enabling technologies, and future directions,'' \emph{IEEE Commun. Surveys \& Tuts.}, vol.~26, no.~1, pp. 196--257, 2024.

\bibitem{alamzadeh2021reconfigurable}
I.~Alamzadeh, G.~C. Alexandropoulos, N.~Shlezinger, and M.~F. Imani, ``A reconfigurable intelligent surface with integrated sensing capability,'' \emph{Scientific Reports}, vol.~11, no.~1, pp. 1--10, 2021.

\bibitem{HRIS_Mag_all}
G.~C. Alexandropoulos, N.~Shlezinger, I.~Alamzadeh, M.~F. Imani, H.~Zhang, and Y.~C. Eldar, ``Hybrid reconfigurable intelligent metasurfaces: {E}nabling simultaneous tunable reflections and sensing for {6G} wireless communications,'' \emph{IEEE Veh. Technol. Mag.}, vol.~19, no.~1, pp. 75--84, 2024.

\bibitem{Eucnc_Terrameta}
S.~Inacio, Y.~Ma, Q.~Luo, L.~Lucci, A.~Kumar, J.~L. Gonzalez~Jimenez, A.~Siligaris, B.~R. amd D.~Mercier, T.~D. Phan, P.~Jack~Soh, S.~Matos, G.~C. Alexandropoulos, L.~M. Pessoa, and A.~Clemente, ``Evaluation of switching technologies for reflective and transmissive {RISs} at sub-{THz} frequencies,'' in \emph{Proc. Joint European Conf. Netw. Commun. \& {6G} Summit}, Poznan, Poland, 2025.

\bibitem{RIS_Scattering_all}
G.~C. Alexandropoulos, N.~Shlezinger, and P.~del Hougne, ``Reconfigurable intelligent surfaces for rich scattering wireless communications: {R}ecent experiments, challenges, and opportunities,'' \emph{IEEE Commun. Mag.}, vol.~59, no.~6, pp. 28--34, 2021.

\bibitem{WavePropTCCN}
G.~C. Alexandropoulos, G.~Lerosey, M.~Debbah, and M.~Fink, ``Reconfigurable intelligent surfaces and metamaterials: {T}he potential of wave propagation control for {6G} wireless communications,'' \emph{IEEE ComSoc TCCN Newslett.}, vol.~6, no.~1, pp. 25--37, Jun. 2020.

\bibitem{ABSence}
C.~Liaskos \emph{et~al.}, ``{ABSense: S}ensing electromagnetic waves on metasurfaces via ambient compilation of full absorption,'' in \emph{Proc. ACM NANOCOM}, Dublin, Ireland, Sep. 2019.

\bibitem{hardware2020icassp}
G.~C. Alexandropoulos and E.~Vlachos, ``A hardware architecture for reconfigurable intelligent surfaces with minimal active elements for explicit channel estimation,'' in \emph{Proc. IEEE ICASSP}, Barcelona, Spain, May 2020, pp. 9175--9179.

\bibitem{10237986}
I.~Alamzadeh, G.~C. Alexandropoulos, and M.~F. Imani, ``Intensity-only omp-based direction estimation for hybrid reconfigurable intelligent surfaces,'' in \emph{Proc. IEEE Int. Symposium Antennas Propag. and USNC-URSI Radio Science Meeting}, Portland, USA, 2023.

\bibitem{sleasman2016microwave}
T.~Sleasman, M.~F. Imani, J.~N. Gollub, and D.~R. Smith, ``Microwave imaging using a disordered cavity with a dynamically tunable impedance surface,'' \emph{Physical Review Applied}, vol.~6, no.~5, p. 054019, 2016.

\bibitem{boyarsky2021electronically}
M.~Boyarsky, T.~Sleasman, M.~F. Imani, J.~N. Gollub, and D.~R. Smith, ``Electronically steered metasurface antenna,'' \emph{Scientific Reports}, vol.~11, no.~1, pp. 1--10, 2021.

\bibitem{RIS_challenges}
G.~C. Alexandropoulos, M.~Crozzoli, D.-T. Phan-Huy, K.~D. Katsanos, H.~Wymeersch, P.~Popovski, P.~Ratajczak, Y.~B{\'e}n{\'e}dic, M.-H. Hamon, S.~Herraiz~Gonzalez, R.~D'Errico, and E.~Calvanese~Strinati, ``{RIS}-enabled smart wireless environments: {D}eployment scenarios, network architecture, bandwidth and area of influence,'' \emph{EURASIP J. Wireless Commun. Netw.}, to appear, 2023.

\bibitem{Alexandropoulos2022Pervasive}
G.~C. Alexandropoulos, K.~Stylianopoulos, C.~Huang, C.~Yuen, M.~Bennis, and M.~Debbah, ``Pervasive machine learning for smart radio environments enabled by reconfigurable intelligent surfaces,'' \emph{Proc. IEEE}, vol. 110, no.~9, pp. 1494--1525, 2022.

\bibitem{9410457}
C.~Huang, Z.~Yang, G.~C. Alexandropoulos, K.~Xiong, L.~Wei, C.~Yuen, Z.~Zhang, and M.~Debbah, ``Multi-hop {RIS}-empowered terahertz communications: A {DRL}-based hybrid beamforming design,'' \emph{IEEE J. Sel. Areas Commun.}, vol.~39, no.~6, pp. 1663--1677, 2021.

\bibitem{RIS_security_AIoT}
A.~Kunz, S.~B.~M. Baskaran, and G.~C. Alexandropoulos, ``Lightweight security for ambient-powered programmable reflections with reconfigurable intelligent surfaces,'' \emph{arXiv preprint:2501.09005}, 2025.

\bibitem{wang2020channel}
Z.~Wang \emph{et~al.}, ``Channel estimation for intelligent reflecting surface assisted multiuser communications: Framework, algorithms, and analysis,'' \emph{{IEEE} Trans. Wireless Commun.}, vol.~19, no.~10, pp. 6607--6620, Oct. 2020.

\bibitem{9726785}
G.~C. Alexandropoulos, I.~Vinieratou, and H.~Wymeersch, ``Localization via multiple reconfigurable intelligent surfaces equipped with single receive {RF} chains,'' \emph{IEEE Wireless Commun. Lett.}, vol.~11, no.~5, pp. 1072--1076, 2022.

\bibitem{10124713}
J.~He, A.~Fakhreddine, C.~Vanwynsberghe, H.~Wymeersch, and G.~C. Alexandropoulos, ``{3D} localization with a single partially-connected receiving {RIS}: Positioning error analysis and algorithmic design,'' \emph{IEEE Trans. Veh. Technol.}, vol.~72, no.~10, pp. 13\,190--13\,202, 2023.

\bibitem{10557771}
R.~Ghazalian, G.~C. Alexandropoulos, G.~Seco-Granados, H.~Wymeersch, and R.~Jäntti, ``Joint {3D} user and {6D} hybrid reconfigurable intelligent surface localization,'' \emph{IEEE Trans. Veh. Technol.}, vol.~73, no.~10, pp. 15\,302--15\,317, 2024.

\bibitem{8313072}
G.~C. Alexandropoulos, ``Position aided beam alignment for millimeter wave backhaul systems with large phased arrays,'' in \emph{Proc. IEEE Int. Workshop Comp. Adv. Multi-Sensor Adaptive Process.}, Curaçao, Dutch Antilles, 2017.

\bibitem{HRIS_ISAC}
I.~Gavras and G.~C. Alexandropoulos, ``Simultaneous communications and sensing with hybrid reconfigurable intelligent surfaces,'' in \emph{Proc. European Conf. Antennas Propag.}, Stockholm, Sweden, 2025.

\bibitem{RFimaging}
K.~Stylianopoulos, P.~Gavriilidis, G.~Gradoni, and G.~C. Alexandropoulos, ``Graph-{CNNs} for {RF} imaging: Learning the electric field integral equations,'' \emph{arXiv preprint arXiv:2503.14439}, 2025.

\bibitem{GGM_ACCESS_2024}
A.~Ghaneizadeh, P.~Gavriilidis, M.~Joodaki, and G.~C. Alexandropoulos, ``Metasurface energy harvesters: State-of-the-art designs and their potential for energy sustainable reconfigurable intelligent surfaces,'' \emph{IEEE Access}, vol.~12, pp. 160\,464--160\,494, 2024.

\bibitem{Duong_relay_selection}
T.~T. Duy, G.~C. Alexandropoulos, T.~T. Vu, N.-S. Vo, and T.~Q. Duong, ``Outage performance of cognitive cooperative networks with relay selection over double-rayleigh fading channels,'' \emph{IET Commun.}, vol.~10, no.~1, pp. 57--67, 2016.

\bibitem{Duong_AF_relaying}
T.~Q. Duong, V.~N.~Q. Bao, H.~Tran, G.~C. Alexandropoulos, , and H.~Zepernick, ``Effect of primary networks on the performance of spectrum sharing {AF} relaying,'' \emph{Electron. Lett.}, vol.~48, no.~1, pp. 25--27, 2012.

\bibitem{DF_relays}
G.~C. Alexandropoulos, A.~Papadogiannis, and P.~C. Sofotasios, ``Outage performance of cognitive cooperative networks with relay selection over double-rayleigh fading channels,'' \emph{HINDAWI J. Computer Netw. Commun.}, vol. 2011, Article ID 560528, 2011.

\bibitem{7997175}
G.~C. Alexandropoulos and M.~Duarte, ``Joint design of multi-tap analog cancellation and digital beamforming for reduced complexity full duplex {MIMO} systems,'' in \emph{Proc. IEEE Int. Conf. Commun.}, Paris, France, 2017.

\bibitem{9933358}
G.~C. Alexandropoulos, M.~A. Islam, and B.~Smida, ``Full-duplex massive multiple-input, multiple-output architectures: Recent advances, applications, and future directions,'' \emph{IEEE Veh. Technol. Mag.}, vol.~17, no.~4, pp. 83--91, 2022.

\bibitem{10447070}
P.~Gavriilidis, I.~Atzeni, and G.~C. Alexandropoulos, ``Metasurface-based receivers with $1$-bit {ADCs} for multi-user uplink communications,'' in \emph{Proc. IEEE Int. Conf. Acoustics, Speech Signal Process}, Seoul, South Korea, 2024, pp. 9141--9145.

\bibitem{zhang2022active}
Z.~Zhang, L.~Dai, X.~Chen, C.~Liu, F.~Yang, R.~Schober, and H.~V. Poor, ``Active {RIS} vs. passive {RIS}: Which will prevail in 6{G}?'' \emph{{IEEE} Trans. Commun.}, vol.~71, no.~3, pp. 1707--1725, 2022.

\bibitem{mine_Active_RIS}
P.~Gavriilidis, D.~Mishra, B.~Smida, E.~Basar, C.~Yuen, and G.~C. Alexandropoulos, ``Active reconfigurable intelligent surfaces: Circuit modeling and reflection amplification optimization,'' \emph{arXiv preprint arXiv:2503.24093}, 2025.

\bibitem{RIS_ISAC_SPM}
S.~P. Chepuri, N.~Shlezinger, F.~Liu, G.~C. Alexandropoulos, S.~Buzzi, and Y.~C. Eldar, ``Integrated sensing and communications with reconfigurable intelligent surfaces: From signal modeling to processing,'' \emph{IEEE Signal Process. Mag.}, vol.~40, no.~6, pp. 41--62, Sep. 2023.

\bibitem{hu2019two}
C.~Hu \emph{et~al.}, ``Two-timescale channel estimation for reconfigurable intelligent surface aided wireless communications,'' \emph{{IEEE} Trans. Commun.}, vol.~69, no.~11, pp. 7736--7747, Nov. 2021.

\bibitem{LZA_TWC_2021}
S.~Lin, B.~Zheng, G.~C. Alexandropoulos, M.~Wen, M.~Di~Renzo, and F.~Chen, ``Reconfigurable intelligent surfaces with reflection pattern modulation: Beamforming design and performance analysis,'' \emph{IEEE Trans. Wireless Commun.}, vol.~20, no.~2, pp. 741--754, 2021.

\bibitem{LHA_TCOM_2021}
L.~Wei, C.~Huang, G.~C. Alexandropoulos, C.~Yuen, Z.~Zhang, and M.~Debbah, ``Channel estimation for {RIS}-empowered multi-user {MISO} wireless communications,'' \emph{IEEE Trans. Commun.}, vol.~69, no.~6, pp. 4144--4157, 2021.

\bibitem{LHG_TWC_2022}
L.~Wei, C.~Huang, Q.~Guo, Z.~Yang, Z.~Zhang, G.~C. Alexandropoulos, M.~Debbah, and C.~Yuen, ``Joint channel estimation and signal recovery for {RIS}-empowered multiuser communications,'' \emph{IEEE Trans. Commun.}, vol.~70, no.~7, pp. 4640--4655, 2022.

\bibitem{BGA_TWC_2024}
M.~Bayraktar, N.~González-Prelcic, G.~C. Alexandropoulos, and H.~Chen, ``Ris-aided joint channel estimation and localization at mmwave under hardware impairments: A dictionary learning-based approach,'' \emph{IEEE Trans. Wireless Commun.}, vol.~23, no.~12, pp. 19\,696--19\,712, 2024.

\bibitem{alexandropoulos2020hardware}
G.~C. Alexandropoulos and E.~Vlachos, ``A hardware architecture for reconfigurable intelligent surfaces with minimal active elements for explicit channel estimation,'' in \emph{Proc. IEEE ICASSP}, Barcelona, Spain, May 2020.

\bibitem{wu2019intelligent}
Q.~Wu and R.~Zhang, ``Intelligent reflecting surface enhanced wireless network via joint active and passive beamforming,'' \emph{{IEEE} Trans. Wireless Commun.}, vol.~18, no.~11, pp. 5394--5409, Nov. 2019.

\bibitem{self_configuring_RIS}
A.~Albanese, F.~Devoti, V.~Sciancalepore, M.~Di~Renzo, and X.~Costa-P{\'e}rez, ``A self-configuring metasurfaces absorption and reflection solution towards {6G},'' in \emph{Proc. IEEE INFOCOM}, May 2022.

\bibitem{DRL_automonous_RIS}
G.~C. Alexandropoulos, K.~Stylianopoulos, C.~Huang, C.~Yuen, M.~Bennis, and M.~Debbah, ``Pervasive machine learning for smart radio environments enabled by reconfigurable intelligent surfaces,'' \emph{Proc. {IEEE}}, vol. 110, no.~9, pp. 1494--1525, 2022.

\bibitem{SORIS}
\BIBentryALTinterwordspacing
E.~Koutsonas, A.-A.~A. Boulogeorgos, G.~C. Alexandropoulos, T.~A. Tsiftsis, and R.~Zhang, ``{SORIS}: A self-organized reconfigurable intelligent surface architecture for wireless communications,'' 2025. [Online]. Available: \url{http://dx.doi.org/10.36227/techrxiv.174439668.85376932/v1}
\BIBentrySTDinterwordspacing

\bibitem{biguesh2006training}
M.~Biguesh and A.~B. Gershman, ``Training-based {MIMO} channel estimation: a study of estimator tradeoffs and optimal training signals,'' \emph{{IEEE} Trans. Signal Process.}, vol.~54, no.~3, pp. 884--893, Mar. 2006.

\bibitem{sengijpta1995fundamentals}
S.~K. Sengijpta, \emph{Fundamentals of statistical signal processing: Estimation theory}.\hskip 1em plus 0.5em minus 0.4em\relax Taylor \& Francis Group, 1995.

\bibitem{HRIS_ISAC_radarconf2025_1}
I.~Gavras and G.~C. Alexandropoulos, ``Tracking-aided multi-user {MIMO} communications with hybrid reconfigurable intelligent surfaces,'' \emph{arXiv preprint arXiv:2504.18846}, 2025.

\bibitem{HRIS_ISAC_radarconf2025_2}
------, ``Communications-centric secure {ISAC} with hybrid reconfigurable intelligent surfaces,'' \emph{arXiv preprint arXiv:2504.20608}, 2025.

\end{thebibliography}

\end{document}